\documentclass[twocolumn]{aastex631}

\usepackage[english]{babel}
\usepackage{bbm}
\usepackage{enumitem}
\usepackage{graphicx}
\usepackage{xcolor}
\usepackage{wrapfig}
\usepackage{soul}
\usepackage{physics}

\definecolor{darkgoldenrod}{rgb}{0.72, 0.53, 0.04}

\usepackage[normalem]{ulem}

\newcommand*{\mcfacts}{\texttt{McFACTS}}
\newcommand*{\pagn}{\texttt{pAGN}}

\newcommand*{\SGfixed}{\texttt{SG\_fixed}}
\newcommand*{\SGscaled}{\texttt{SG\_scaled}}

\newcommand*{\BinaryParameters}{\vec{\lambda}}

\newcommand*{\chieff}{\chi_{\mathrm{eff}}}
\newcommand*{\redshift}{z}
\newcommand*{\metallicity}{\mathcal{Z}}

\newcommand*{\msun}{M_{\odot}}

\newcommand*{\WeightVolumetric}{\mathcal{R}}
\newcommand*{\WeightIntrinsic}{\rho}
\newcommand*{\WeightDetection}{r}
\newcommand*{\RateIntrinsic}{\mu_{s}}
\newcommand*{\RateDetection}{\mu}

\newcommand*{\citeOSG}{\citet{osg07,osg09}}

\newcommand{\AffiliationCCRG}{
  Center for Computational Relativity and Gravitation, 
  Rochester Institute of Technology, 
  Rochester, New York 14623, USA 
}
\newcommand{\AffiliationGSFC}{
  Gravitational Astrophysics Laboratory, 
  NASA Goddard Space Flight Center, 
  Greenbelt, MD 20771, USA 
}
\newcommand{\AffiliationCCA}{
  Center for Computational Astrophysics,
  Flatiron Institute,
  New York, NY 10010, USA
}
\newcommand{\AffiliationAMNH}{
  Department of Astrophysics, 
  American Museum of Natural History, 
  New York, NY 10024, USA 
}
\newcommand{\AffiliationBMCC}{
 Department of Science,
 BMCC, City University of New York,
 New York, NY 10007, USA
}
\newcommand{\AffiliationNMSU}{
  New Mexico State University, 
  Department of Astronomy, 
  PO Box 30001 MSC 4500, 
  Las Cruces, NM 88001, USA 
}
\newcommand{\AffiliationCityUniversity}{
  Graduate Center, 
  City University of New York, 
  365 5th Avenue, 
  New York, NY 10016, USA 
}

\begin{document}
\title{
       \mcfacts{} III: Compact binary mergers from AGN disks over an entire synthetic universe
}
\author[0000-0001-7099-765X]{Vera Delfavero}
\affiliation{\AffiliationGSFC}

\author[0000-0002-5956-851X]{K. E. Saavik Ford}
\affiliation{\AffiliationBMCC}
\affiliation{\AffiliationAMNH}
\affiliation{\AffiliationCityUniversity}
\affiliation{\AffiliationCCA}

\author[0000-0002-9726-0508]{Barry McKernan}
\affiliation{\AffiliationBMCC}
\affiliation{\AffiliationAMNH}
\affiliation{\AffiliationCityUniversity}
\affiliation{\AffiliationCCA}

\author[0000-0001-7163-8712]{Harrison E. Cook}
\affiliation{\AffiliationNMSU}

\author[0000-0003-2430-9515]{Kaila Nathaniel}
\affiliation{\AffiliationCCRG}

\author[0000-0003-0738-8186]{Jake Postiglione}
\affiliation{\AffiliationAMNH}
\affiliation{\AffiliationCityUniversity}

\author[0009-0005-5038-3171]{Shawn Ray}
\affiliation{\AffiliationAMNH}
\affiliation{\AffiliationCityUniversity}

\author[0009-0008-5622-6857]{Emily McPike}
\affiliation{\AffiliationAMNH}
\affiliation{\AffiliationCityUniversity}

\author[0000-0001-5832-8517]{Richard O'Shaughnessy}
\affiliation{\AffiliationCCRG}

\date{\today}

\clearpage{}\begin{abstract}

The Active Galactic Nuclei (AGN) channel for the formation of binary 
    black hole (BBH) mergers has been previously studied as a potential formation channel
    for the merging compact binaries observed by the LIGO/Virgo/KAGRA (LVK)
    scientific collaboration.
The first two papers in this series explored the \mcfacts{} code
    for the evolution of black hole orbits in AGN accretion disks for individual
    galaxy models and described the characteristics of predicted BBH populations
    in realizations of those models
    (such as the correlation between mass ratio and aligned spin).
In this work, we explore the impact of the properties of AGN host galaxies
    and assume an AGN lifetime and cosmological model for the density of
    AGN in a universe like our own.
By sampling from an inferred population of AGN,
    we marginalize over galaxy mass to predict a population of 
    BBH mergers observable by modern ground-based gravitational wave observatories.
We find that for reasonable assumptions,
    AGN disk environments may account for massive BBH mergers
    such as GW190521 and GW190929\_012149.
We find that the majority of observable BBH mergers from our simulation
    are expected to originate in galaxies with a super-massive black hole
    between $10^7 \msun$ and $10^{9.4} \msun$.
We also find that if hierarchical mergers from AGN disks
    account for a substantial part of the LVK population,
    our current models require an AGN lifetime of $0.5 - 2.5$ Myr.

\end{abstract}
\clearpage{}

\section{Introduction}
\label{sec:intro}
In the past decade, the LIGO-Virgo-KAGRA (LVK) collaboration
    has reported gravitational wave observations from about 90
    compact binary mergers, of which the majority are suspected binary black hole
    (BBH) mergers
    \citep{GWTC-1, GWTC-2, GWTC-3, GWTC-2p1}.
The progenitors of BBH mergers observed this way
    can form in a number of different channels:
    both by the isolated evolution of massive binary stars
    \citep{broekgaarden2021formation, broekgaarden2021impact,
    st_inference_interp, Belczynski2020-EvolutionaryRoads,
    Zevin2020,Olejak2020,Stevenson2015popsyn,2020arXiv201016333B,
    posydon, Stevenson2022, COSMICZevin2021, COSMICWong2022,Olejak2024Spin}
    and by dynamical encounters in dense stellar environments such
    as globular clusters
    \citep{gwastro-popsynVclusters-Rodriguez2016,
    Morscher_2015,Miller_2002,Portegies_Zwart_2004,
    DiCarlo_2020lfa,Wang_2022,Rapster2024,Li2024Origins,Li2024Field}
    and Active Galactic Nuclei (AGN) accretion disks
    \citep{McK14,Bartos17,Stone17,ArcaSedda23,Yang19,Gayathri2022}.

\begin{figure*}[t]
\centering
\includegraphics[width=\textwidth]{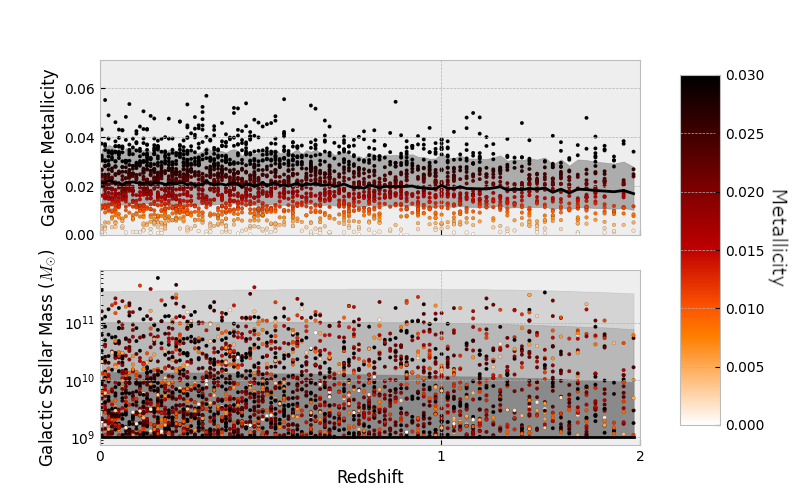}
\caption{\label{fig:orange}
    \textbf{Sampling high mass galaxies in the local Universe:}
    (top): Galactic metallicity of galaxy samples in each
        100 Myr epoch (colored by metallicity).
    The black line is a mass-weighted mean of metallicity,
        and the shaded region illustrates 0.5 dex around the average.
    (bottom): Galactic stellar mass of galaxy samples
        (colored again by metallicity).
    Shaded regions enclose 0.68, 0.95, and 0.997 percent of samples
        above the $10^9 \msun$ threshold.
    One in every hundred samples is shown
        for visualization.
}
\end{figure*}

The AGN formation channel is characterized by both an efficient rate of BBH formation 
    due to close encounters in AGN disks \citep{Bellovary16,Secunda19,Whitehead24} and 
    a high rate of retention of merged BHs (post merger-kick)
    in a deep potential well \citep{Ford22}, 
    leading to several overall expectations for the BBH mergers from AGN.
These expectations can include:
 (i) a high rate of hierarchical mergers, including InterMediate Mass Black Hole (IMBH) formation \citep{McK12,Yang19,Tagawa21}
    relative to other channels, 
 (ii) a bias away from equal-mass mergers ($q=M_{2}/M_{1}<1$) \citep{Secunda20,Tagawa20,McK20a},
 (iii) a bias towards $\chi_{\rm eff}>0$ due to spin torquing towards alignment with the AGN gas disk \citep{Tagawa20b}, and
(iv) a possible bias towards eccentric (including highly eccentric) mergers \citep{Samsing22,Calcino24}, 
(v) possible ($q,\chi_{\rm eff}$) anti-correlation \citep{qX22,Santini23}.
 An understanding of the fractional contribution of the AGN channel to the observed BBH merger rate 
    will allow us to restrict several poorly
    constrained AGN disk parameters.
These include average disk density, lifetime, size and duty cycle \citep{Vajpeyi22}.
These also may include nuclear star cluster (NSC) parameters
    such as initial BH mass function,
    radial density distribution,
    and the overall population of stellar BHs and IMBHs
    in galactic nuclei \citep{McK18}.

 In \citet{mcfactsI}, we introduced the new, public, open-source code \mcfacts{} (Monte Carlo
 For AGN Channel Testing and Simulation) and demonstrated several parameter space tests that can be carried out by the code.
 In \citet{mcfactsII}, we used \mcfacts{} to carry out a study of ($q,\chi_{\rm eff}$) parameter space
 as a function of disk model and several dependent parameters in individual galaxy models.
In this work, we use \mcfacts{} to study the intrinsic and observable
    BBH merger population from a synthetic universe of AGN,
    drawn from a set of cosmological assumptions
    (e.g. galactic stellar mass function, AGN density).
In doing so, we sample from sets of \mcfacts{} simulations
    for galaxy models which scale NSC mass
    and super-massive black hole mass with galaxy mass.
We assume AGN disks take the form of the commonly-used model \citep{SG03},
    and investigate the BBH merger populations that result.
\citet{mcfactsII} explores differences between \citet{SG03} and
    \citet{TQM05} for standard disk parameterization, 
    finding better agreement with LVK observations for \citet{SG03}.

 \section{Methods}
\label{sec:methods}
In this section, we explore our assumed constraints on AGN host galaxies,
    and how we enforce those 
    on AGN populations in realistic galaxies, and how we enforce those
    constraints to study their BBH merger populations with \mcfacts{}.
We explore our predictions for the BBH population of 
    galaxies with $M_* \in [10^9, 10^{13}]$,
    and explore a realistic sample of galaxies in 
a synthetic universe like our own.
We further explore our methods for estimating merger
    rates and populations from the AGN channel,
    first in discrete regions of fixed volume in the synthetic universe
    (where galaxy populations are relatively constant),
    and then in a larger volume (out to redshift 2).
Finally, we predict the LVK BBH detectable population in that larger volume,
    as well as the contribution of AGN channel BBH mergers to
    the overall LVK detection rate.

\begin{figure*}[t]
\centering
\includegraphics[width=3.375 in]{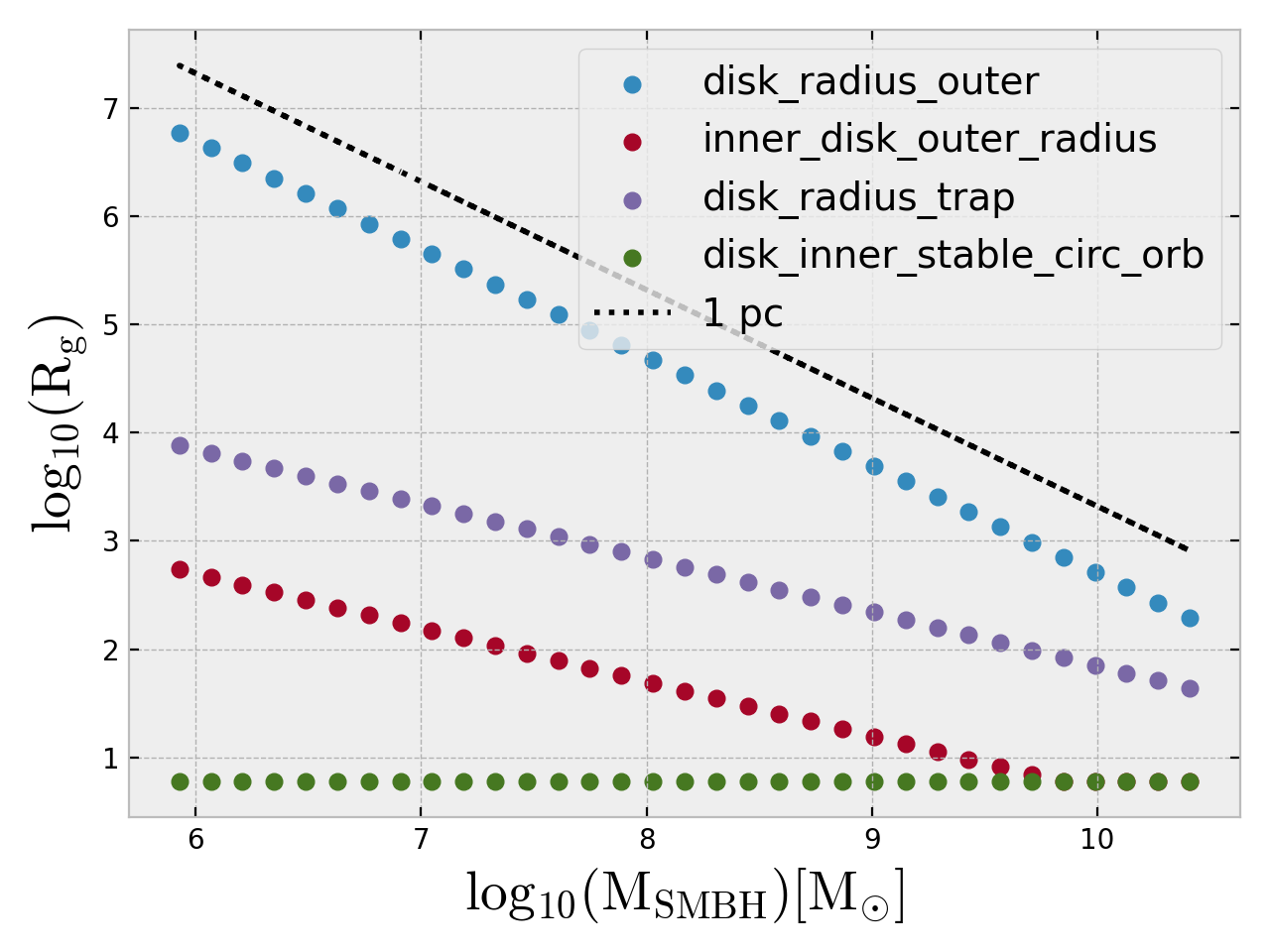}
\includegraphics[width=3.375 in]{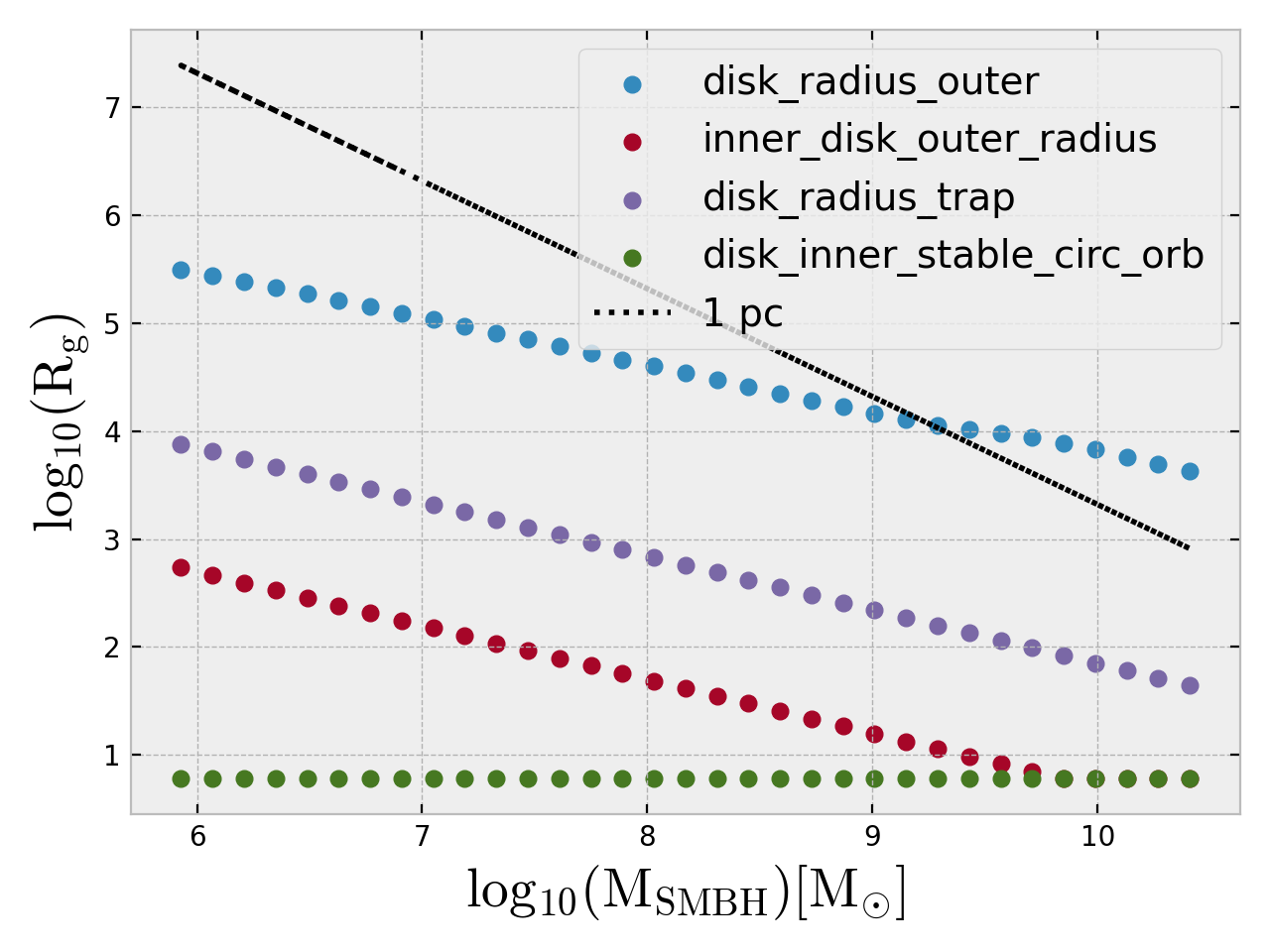}
\caption{\label{fig:disk}
    \textbf{Disk properties as a function of SMBH mass:}
    (left) Disk properties for the Sirko-Goodman disk with a fixed outer disk radius
        of $0.25 \mathrm{pc}$.
    (right) Disk properties for the Sirko-Goodman disk with outer disk radius
        determined by opacity, using \pagn{}.
    This figure shows the outer disk radius, trap radius, ISCO radius,
        and outer radius of the inner disk
        for galaxies of different SMBH mass.
    The ISCO radius is always 6 $R_g$.
}
\end{figure*}

\subsection{AGN populations in realistic galaxies}
\label{sec:postproc}
Effectively sampling a realistic population of AGN by sampling galaxy
    properties in the Universe
    requires broad assumptions about AGN and the history of the Universe.
In order to estimate the properties of the AGN in a universe like our own,
    we sample galactic populations from the \citet{fontana2006}
    Galactic Stellar Mass Function (GSMF)
    and \citet{Madau2017} (Eq. 6) metallicity evolution
    assumptions.
Because galactic populations change throughout the history of the Universe,
    we sample galactic populations which represent 
    each 100 Myr epoch of the evolutionary history of the Universe.
These assumptions have been used extensively to sample star formation
    for predicting gravitational wave progenitors through the isolated binary
    evolution formation channel
    \citep{Belczynski2020-EvolutionaryRoads,
        st_inference_interp}.

Figure \ref{fig:orange} visualizes the impact of our choices for
    the GSMF and metallicity distribution on galaxy sample populations.
In this work, we consider only galaxies with a stellar mass greater than
    $10^9 \mathrm{M}_{\odot}$, which should have a minimal impact on predicted BBH populations,
    given the low rate of AGN in such low mass galaxies
    (relative to higher mass galaxies, see \citet{Reines2013}),
    as well as the declining contribution to the overall BBH merger rate at lower galaxy/SMBH mass, as we will show below.
We sample 10,000 galaxies in each epoch,
    many of which fall below this threshold.
We also restrict ourselves to AGN activity in the local universe
    (out to redshift 2).We note that our choice of galactic metallicity distribution characterizes galaxies
    according to observations at different redshifts
    \citep{Madau2017},
    rather than by galactic stellar mass
    \citep[E.g.][]{Ma15,Nakajima_2023}.
The reason for this choice is that 
    we only use the metallicity to determine galaxy morphology
    (in conjunction with galactic stellar mass),
    and we don't want to use a galactic metallicity distribution
    which may make late-type (spiral) galaxies irrelevant
    when doing so may neglect the uncertainties introduced when using a simple power law
    to estimate the SMBH mass of a galaxy.
For a qualitative example, the average (mass-weighted) metallicity of
    galaxy samples above a galactic stellar mass of $1.7 \times 10^{10} \msun$
    is greater than $\metallicity = 0.03$
    when assuming a single Schecter function consistent with \citep{Furlong2015} and
    the \citep{Ma15} metallicity relation.
We will find later, in Figure \ref{fig:bbh-bins} that this corresponds to the minimum
    galactic stellar mass to have a an average of at least one BBH merger in an AGN disk
    during a 10 Myr lifetime.
Using such a metallicity to determine the morphology of galaxy samples
    may introduce an undue bias toward an elliptical morphology.

Early- and late-type galaxies also have different star formation environments,
    such as the metallicity of star formation.
The authors of \citet{Peng2015} have characterized the dependence of star forming metallicity on
    the stellar mass of a galaxy (for early- and late-type galaxies).
In order to classify galaxy samples based on their stellar mass and metallicity,
    we fit the data from Figure 2a of that paper (kindly provided by the authors)
    with a Gaussian process.
Using this model, we assign
    the morphology of
    each galaxy sample as either early- 
    or late-type by estimating which branch the sample is closest to.

Two characteristics that determine the properties of AGN in a host galaxy
    are their Super-Massive Black Hole (SMBH) mass
    and Nuclear Star Cluster (NSC) mass.
A two-parameter model such as this is insufficient for describing AGN disk environments 
    in all types of galaxies throughout our Universe.
However, in this first study of compact binary mergers from AGN with properties
    sampled from a GSMF, we make this simplification.
The SMBH mass of a galaxy can be approximated by a power law fit to its stellar mass
    \citep{Schramm_2013}:
\begin{equation}\label{eq:smbh-mass}
M_{\mathrm{SMBH}} \propto 7.07 \times 10^{-5} M_*^{1.12}.
\end{equation}
Regardless of AGN activity, 
    all galaxies are assumed to have a NSC. 
Despite the 
    previously studied decline
    in NSC occupation fraction at large SMBH mass
\citep[E.g.][]{Neumayer2020},
    processes that produce NSC continue to operate at large SMBH mass 
    (e.g. dynamical friction and nuclear star formation),
    while the detectability of NSCs of similar size as found around low mass SMBH also declines. 
Thus we assume a plateau in mass of our NSCs at $10^8 \msun$, but do not assume their absence.

In this work, we assume there are two 
    galaxy morphologies:
    passive (early-type) and star-forming (late-type) galaxies.
The mass of the NSC can similarly be approximated by a power law fit to a galaxy's stellar mass
    (albeit one power law for star-forming/late-type galaxies, and another for passive/early-type galaxies)
    \citep{Neumayer2020}:
\begin{eqnarray}\label{eq:nsc-mass}
\mathrm{log}_{10} M_{\mathrm{NSC}, \mathrm{early}} = 0.48 \mathrm{log}_{10} (\frac{M_*}{10^{9} M_{\odot}}) + 6.51 \\ \nonumber
\mathrm{log}_{10} M_{\mathrm{NSC}, \mathrm{late}} = 0.92 \mathrm{log}_{10} (\frac{M_*}{10^{9} M_{\odot}}) + 6.13,
\end{eqnarray}
subject to the constraint of a maximum mass of $10^8 \msun$, as explained above.

In order to estimate the BBH merger rate in our volume,
    we must normalize the number of AGN which begin an episode of accretion
    in each 100 Myr epoch to the AGN number Density (AGND) of the Universe.
In this work, we make the simplifying assumption that the number density of 
    AGN is proportional to the density of underlying galaxies (i.e. there is no
    bias in having AGN from galaxies of different masses or metallicities).
We define the duty cycle of an AGN as its lifetime ($\mathrm{T}_{\mathrm{lifetime}}$)
    divided by the epoch duration ($\mathrm{dT}_{\mathrm{epoch}}$):
$\mathrm{D}_{\mathrm{AGN}} = \mathrm{T}_{\mathrm{lifetime}} / \mathrm{dT}_{\mathrm{epoch}}$.
The number density of AGN which begin an episode of accretion per year is therefore
dependent on our assumed AGN lifetime:
\begin{equation}\label{eq:duty-cycle}
\frac{N_{\mathrm{AGN}}}{\mathrm{Mpc}^{3}\mathrm{yr}} = \Phi(z) * \frac{1}{
    \mathrm{D}_{\mathrm{AGN}}\mathrm{dT}_{\mathrm{epoch}}},
     = \frac{\Phi(z)}{\mathrm{T}_{\mathrm{lifetime}}},
\end{equation}
    where $\Phi(z)$ is the AGND at a specified redshift.

Observations in different wavelengths can be used to infer
    an AGND \citep{Ueda2014,Lyon2024}.
We
    adopt an AGND inferred from infrared bolometric luminosity
    estimates \citep{Lyon2024},
    integrating over a single luminosity bin for $\log_{10}(L/L_{\odot}) \in [10.5,11]$
    to estimate the total number of AGN
    in a $\mathrm{Mpc^{3}}$ volume of the Universe
    for a given redshift.
This is consistent with the $\log_{10}(L/L_{\odot}) = 10.75$ line in
    Figure 7 of \citet{Lyon2024}.

Finally, we have a population of early- and late-type galaxy samples with 
    $M_*$, $M_{\mathrm{SMBH}}$, and $M_{\mathrm{NSC}}$ which represent
    the AGN population beginning a period of accretion in a $\mathrm{Mpc}^3$ volume
    for each 100 Myr epoch of evolutionary history.
We further separate these galaxy samples into bins according to $M_*$,
    allowing us to characterize the AGN for a synthetic universe.
The total AGND for a given epoch (Eq. \ref{eq:duty-cycle})
    is divided among these types and mass bins (according to GSMF samples)
    in order to properly weigh
    sample mergers from each \mcfacts{} simulation
    when integrating over evolutionary history.
As there are $N=100$ \mcfacts{} galaxy realizations in each mass bin,
    the weight assigned to each BBH merger (indexed by $k$)
    in a given mass bin, for a given galaxy morphology, in a particular epoch
    of evolutionary history is:
\begin{equation}\label{eq:sample-weights}
\WeightVolumetric{}_k = \frac{\Phi(z)}{\mathrm{T}_{\mathrm{lifetime}}}\frac{1}{N}f_{\mathrm{GSMF}},
\end{equation}
    where $f_{\mathrm{GSMF}}$ is the fraction of GSMF samples in a particular mass bin
    (of early or late type).

Combining these mass and type sub-populations for a given epoch of
    evolutionary history yields a set of samples (with parameters $\BinaryParameters$) 
    which describe
    the density ($\WeightVolumetric(\BinaryParameters)$) of BBH mergers in each epoch
    (in units of $\mathrm{Mpc}^{-3}\mathrm{yr}^{-1}$).
While this $\WeightVolumetric$ is an ``intrinsic'' population of merger progenitors,
    for the rest of this work,
    we refer to the population of mergers integrated over evolutionary history
    as the ``intrinsic'' population.
Classically, that integral is carried out by weighing the merger density ($\WeightVolumetric$)
    by the differential comoving volume ($\mathrm{d}V_c/\mathrm{d\redshift}$),
    where $\redshift$ is the redshift of a given merger.
This redshift is different from the redshift of formation
    by the consideration of some delay time
    (in this case, the time between the beginning of an AGN cycle and a
    particular merger event).
As we assume isotropy, our differential comoving volume
    represents the volume of a thin spherical shell 
    (and therefore includes a factor of $4\pi$ to account for
    the integral over solid angle).
Throughout this work, we adopt the \texttt{Planck2015} cosmology
    \citep{Planck2015} for consistency with prior work,
    implemented through the astropy software package
    \citep{astropy:2013,astropy:2018,astropy:2022}.

We carry out the integral in cosmological history over the discrete set of
    our samples ($k$) in our whole synthetic universe:
\begin{equation}\label{eq:weight-intrinsic}
\RateIntrinsic = \sum\limits_k \WeightIntrinsic_k = 
    \sum\limits_k \WeightVolumetric_k
    \frac{\mathrm{d}V_c}{\mathrm{d}\redshift} (\redshift_k)
    \frac{\mathrm{d}\redshift}{\mathrm{d}t_m} (\redshift_k)
    \frac{\mathrm{d}t_m}{\mathrm{d}t_{\mathrm{det}}} (\redshift_k)
    \mathrm{d}t \; ,
\end{equation}
where $\frac{\mathrm{d}\redshift}{\mathrm{d}t_m}$ changes coordinates from
    redshift (of merger) to time (of merger),
    $\frac{\mathrm{d}t_m}{\mathrm{d}t_{\mathrm{det}}}$ changes coordinates from
    time (of merger) to detector time,
    and $\mathrm{d}t$ is the duration of one cosmological epoch
    ($100 \mathrm{Myr}$).
These sample weights ($\WeightIntrinsic_k$) are used throughout the rest of this paper
    to describe the intrinsic population of mergers as a density
    ($\WeightIntrinsic(\BinaryParameters)$), and have units of $[\mathrm{yr}^{-1}]$.
$\RateIntrinsic$ represents the predicted number of BBH mergers per year
    in a synthetic universe in a volume out to the maximum simulated redshift.

\subsection{Detection Model}
\label{sec:detection}
We employ a single-detector sensitivity model 
    \citep[see Section II.B.1 of][]{st_inference_interp}
    to model an LVK detector sensitivity consistent with the third LVK observing run.
These methods are similar to prior work by others \citep{DominikIII}.
For our sample mergers, the difference between the intrinsic rate per sample
    ($\WeightIntrinsic_k$) and detection rate per sample ($\WeightDetection_k$) appears simply as 
    $\WeightDetection_k = \WeightIntrinsic_k p_{\mathrm{det}}(\BinaryParameters_k)$.

The detection probability, $p_{\mathrm{det}}$, is estimated using the following
    model choices \citep[Consistent with][]{st_inference_interp}:
First, the optimal matched-filter SNR for a single-detector is interpolated,
    using training data calculated with \texttt{lalsuite} \citep{lalsuite},
    for the
    \texttt{IMRPhenomPv2} phenomenological waveform
    \citep{IMRPhenomPv2},
    and the \texttt{SimNoisePSDaLIGOaLIGO140MpcT1800545}
    Point Spread Distribution \citep{P1200087} (with a $140 \mathrm{Mpc}$ BNS range).
Second, the probability of detection is interpolated from data available
    at \url{https://pages.jh.edu/˜eberti2/research/}
    \citep{DominikIII},
    with an SNR threshold of 8.
Finally, a correction factor is introduced for Volume/Time estimates
    in O3 \citep{T1800427}, consistent with
    \citet{LIGO-O3-O3b-RP}.

Integrating this density over the parameter-space of individual binaries
    ($\BinaryParameters$),
    we estimate the total number of expected gravitational wave observations
    from BBH mergers in AGN disks for our assumed synthetic universe, $\RateDetection$:
\begin{equation}\label{eq:detection-rate}
    \RateDetection = T_{\mathrm{obs}} \sum\limits_k \WeightDetection_k = T_{\mathrm{obs}} \sum\limits_k \WeightIntrinsic_k p_{\mathrm{det}}(\BinaryParameters_k)
\end{equation}
where $T_{\mathrm{obs}}$ is the time during the third observing run
    when at least two detectors were observing.
\footnote{
The reasons for using a single-detector sensitivity model and a two-detector observing time
    is that in practice, many events observed only in one detector are dismissed as insignificant
    or marginal detections.
Therefore a single-detector observing time would be optimistic.
}
As thus far, no confident LVK observations have claimed to have
    a total mass greater than $200 \msun$, we do not consider
    the detection of mergers with total mass greater than that limit.

\subsection{Properties of the disk for non-standard SMBH mass}
\label{sec:disk}

Figure \ref{fig:disk} exhibits how the assumed properties
    of our AGN disk model depend on an assumed SMBH mass.
Among them, the outer disk radius ($R_{\mathrm{outer}}$),
    trap radius ($R_{\mathrm{trap}}$),
    and the outer radius of the inner disk ($R_{\mathrm{inner}}$)
    are defined in the \mcfacts{}  configuration in terms of 
    $R_g = G * M_{\mathrm{SMBH}} / c^2$.
The standard assumptions for these quantities are:
    $R_{\mathrm{outer}} = 50,000 R_g$,
    $R_{\mathrm{trap}} = 700 R_g$,
    and
    $R_{\mathrm{inner}} = 50 R_g$ around a $M_{\mathrm{SMBH}} = 10^8 \msun$ SMBH.
In such a system, $R_{\mathrm{outer}} = 50,000 R_g$ corresponds to
    a distance of $\sim 0.25$ pc, which is an appropriate outer radius,
    given traditional theoretical considerations on the sizes of
    accretion disks ($\order{10^4} R_g$),
    coupled with observational evidence that real disks are factors of a few times larger
    \citep[e.g.][]{Morgan2010}.
Our value for the location of a migration trap corresponds to the location
    of the migration trap found by \citet{Bellovary16} for the
    \citet{SG03} disk model of $M_{\mathrm{SMBH}} = 10^8 \msun$.
The separation of inner disk objects at $R_{\mathrm{inner}} = 50 R_g$
    is due to the decay time of a $10 \msun$ black hole
    in a circular orbit merging with the SMBH
    ($\sim 38$ Myr;
    i.e. the same order of magnitude as a long AGN disk lifetime).
In this work, we scale $R_{\mathrm{inner}}$ and $R_{\mathrm{trap}}$
    by $\sqrt{M_{\mathrm{SMBH}}}$ (see Figure \ref{fig:disk}).
This scaling holds a point mass at a constant $t_{\mathrm{GW}}$ from
    the center of the SMBH,
    where $t_{\mathrm{GW}}$ is the time it would take for the point mass
    to inspiral by gravitational wave radiation
    \citep{Peters1964}.
For the most massive galaxies, $R_{\mathrm{inner}}$ predicted from this 
    relation can
    decrease to less than the Innermost Stable Circular Orbit (ISCO).
In this case, $R_{\mathrm{inner}}$ is set to $R_{\mathrm{ISCO}}$
    for the simulation.
These choices allow us to generate SG disks which solve disk equations
    for different choices of $M_{\mathrm{SMBH}}$.

We explore two sets of disk models (both Sirko-Goodman):
    one model (henceforth \SGfixed{})
    using the internal SG disk profile in \mcfacts{},
    with a fixed outer disk radius of 0.25 pc
    and a fixed constant opacity;
    the other model using \pagn{} \citep{pAGN}
    (henceforth \SGscaled) to interpolate opacity 
    as a function of disk radius.
The software package \pagn{} \citep{pAGN}
    generates AGN disk profiles with customizable properties.
For the \SGscaled{} disk models, we set the outer disk radius to the location
    where the opacity drops to or below
    the value at the innermost radius of the disk model.
The \SGfixed{} disk model uses a constant opacity of $0.575 \mathrm{cm}^2\mathrm{g}^{-1}$,
    which is higher in the outer disk than the \SGscaled model.
We use \pagn{} profiles for other quantities as well:
    The sound speed of gas in the disk,
    the density of the disk (different from surface density),
    the pressure and temperature gradient of the disk,
    and orbital frequency.
For the \SGfixed{} disk model, we interpolate a pre-recorded \pagn{} estimate
    of these profiles for the fiducial disk model and interpret these
    estimates for galaxies of various SMBH sizes by scaling with $R_g$.
The assumptions underlying each disk model
    are described by the p3\_SG\_fixed.ini and 
    p3\_SG\_scaled.ini configuration files.

 \section{Results}
\label{sec:results}
\subsection{Number of BBH mergers as a function of galaxy mass}
\begin{figure*}
  \centering
  \includegraphics[width=3.375 in]{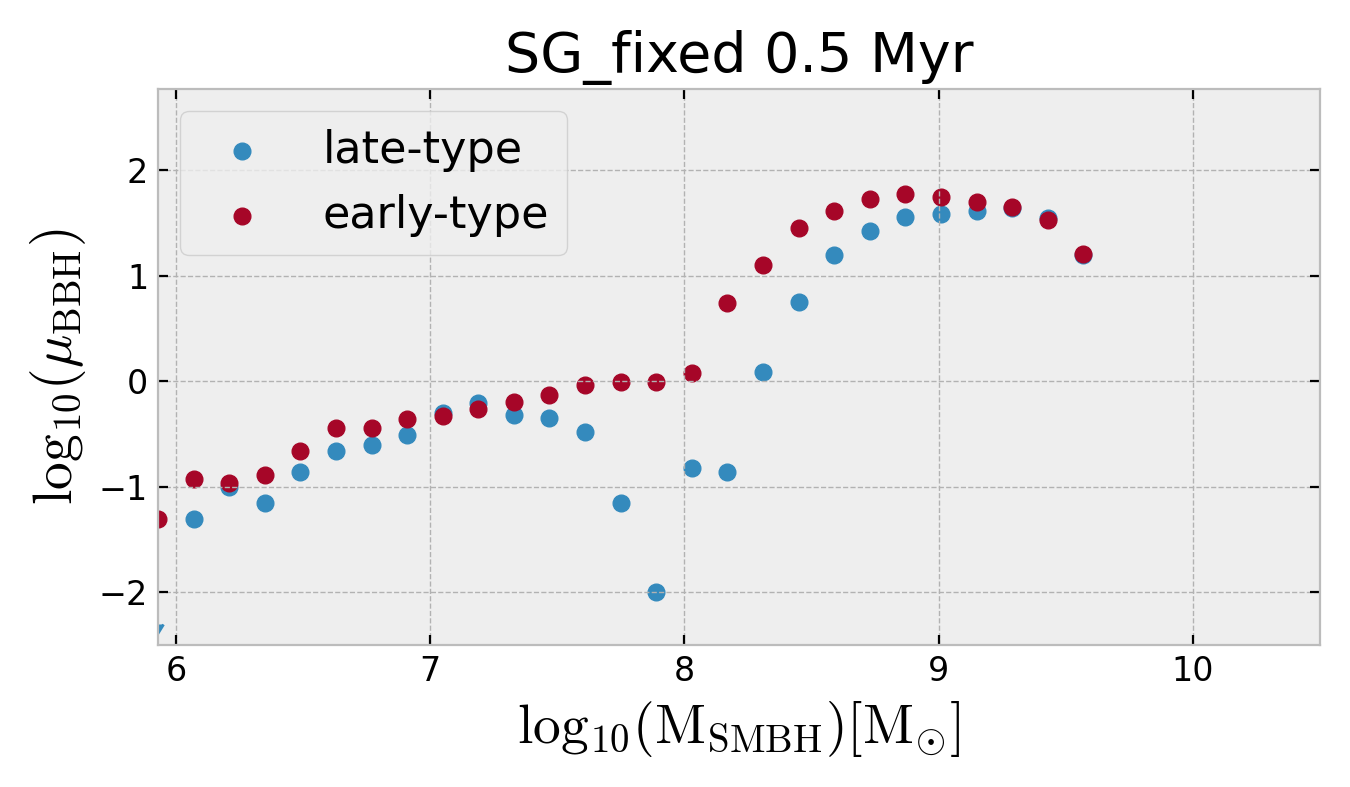}
  \includegraphics[width=3.375 in]{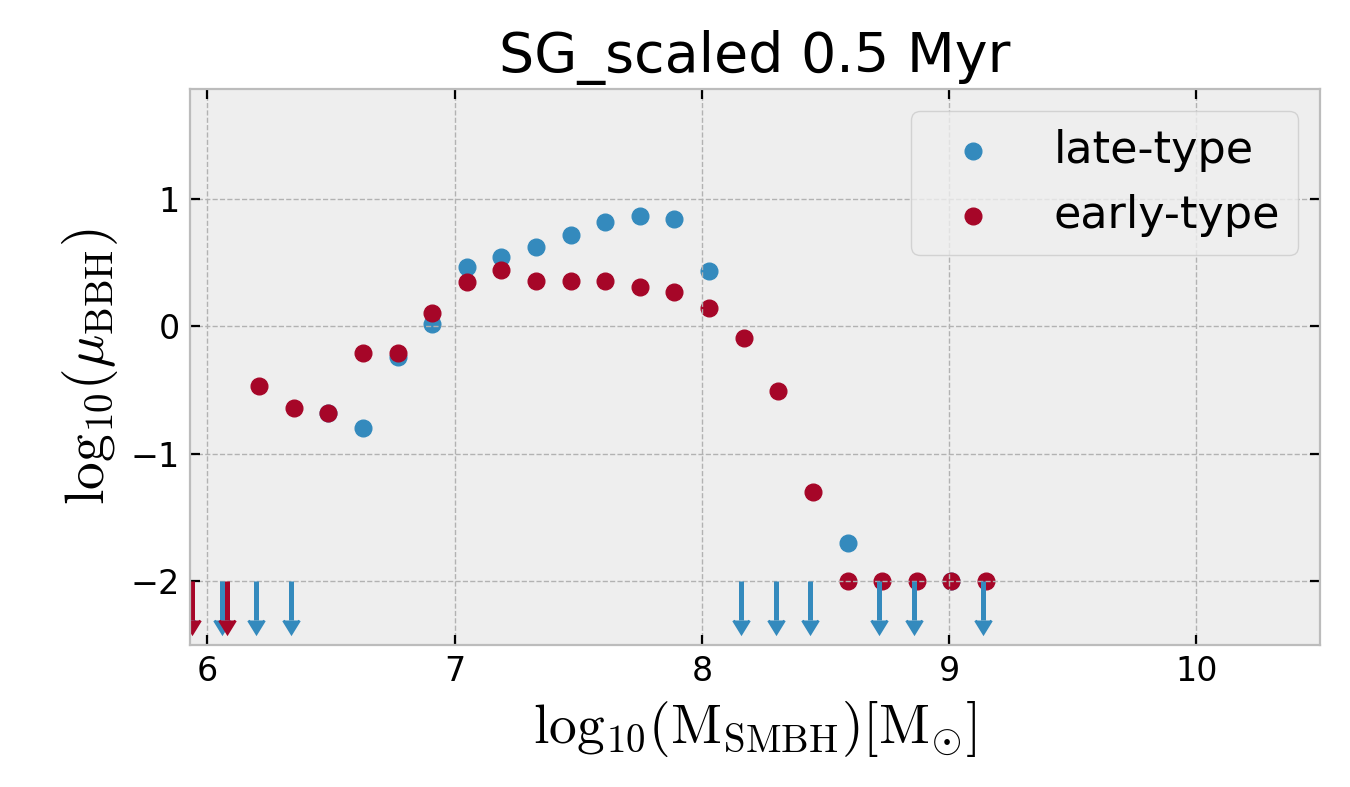} \\
  \includegraphics[width=3.375 in]{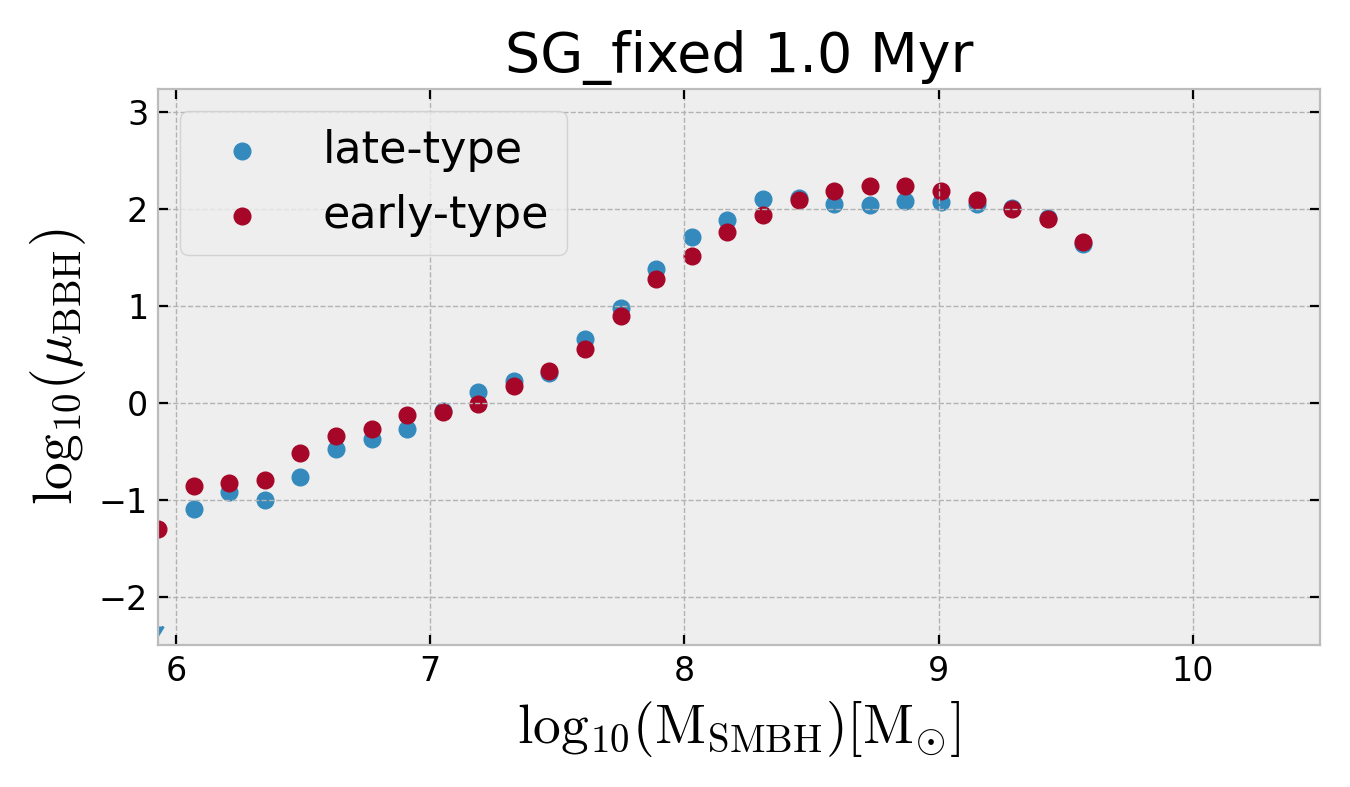}
  \includegraphics[width=3.375 in]{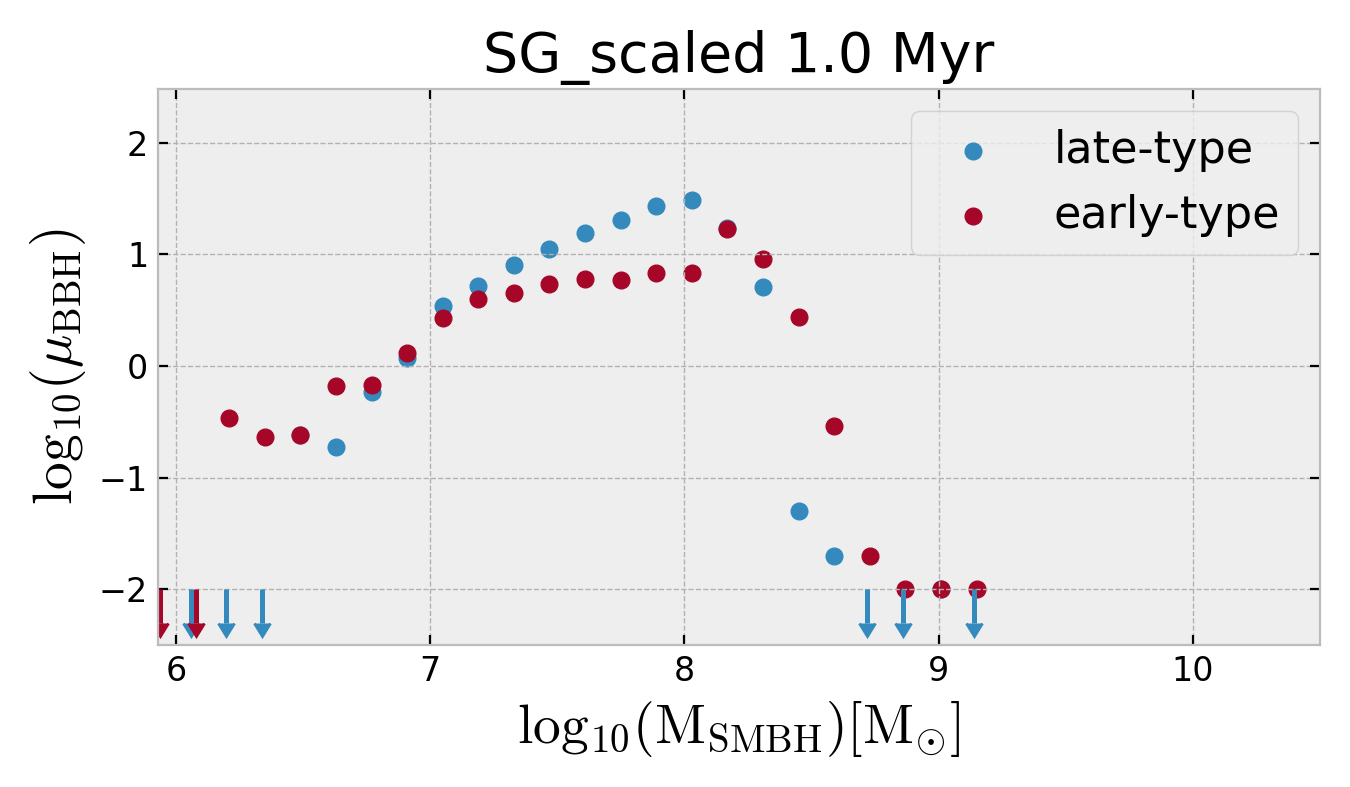} \\
  \includegraphics[width=3.375 in]{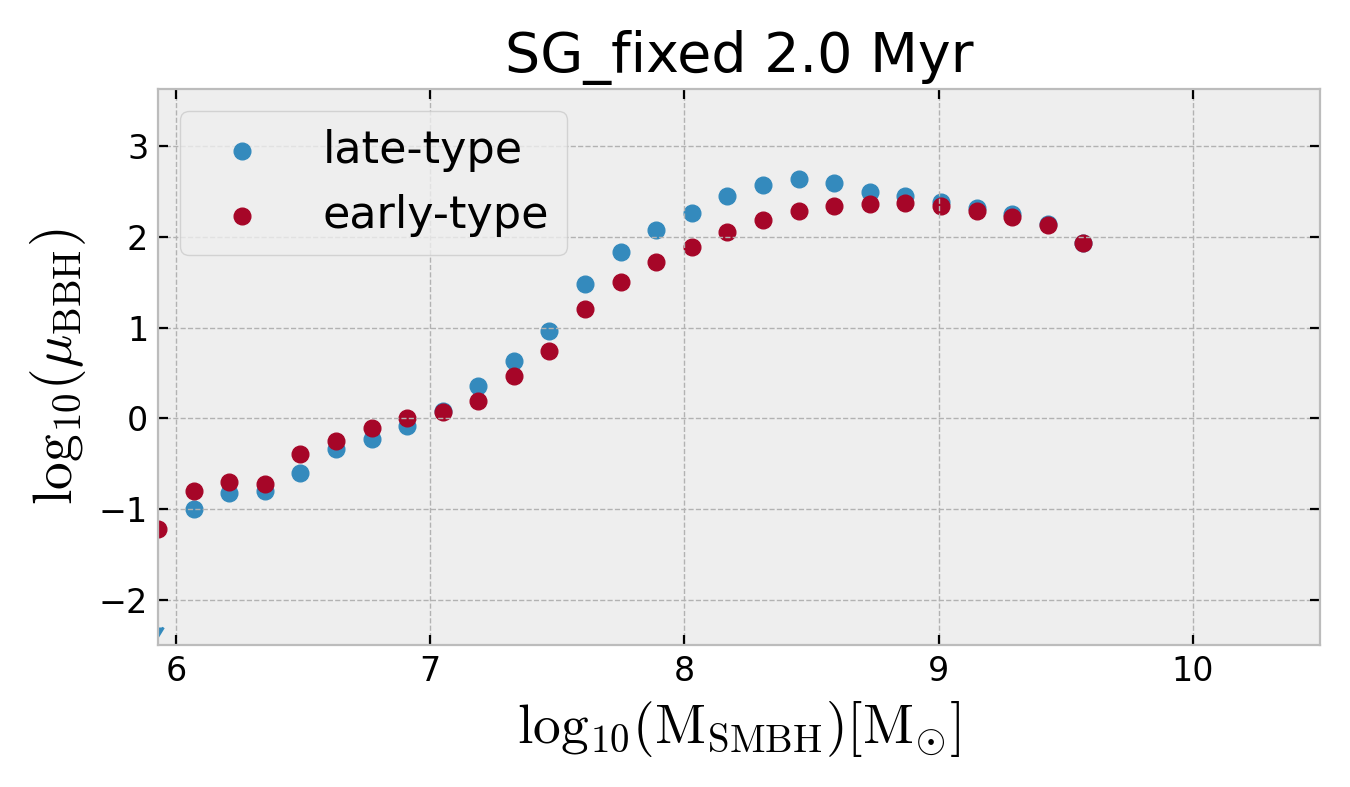}
  \includegraphics[width=3.375 in]{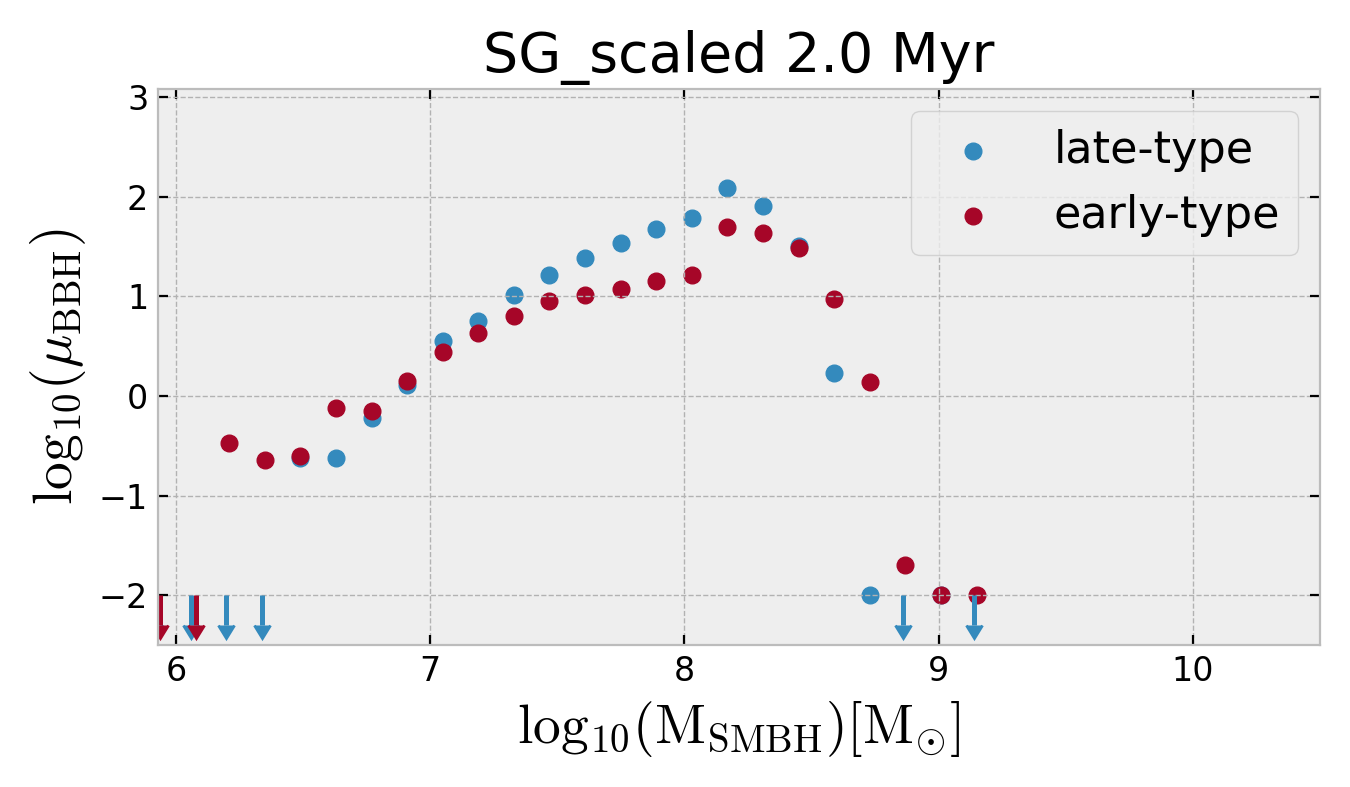} \\
  \includegraphics[width=3.375 in]{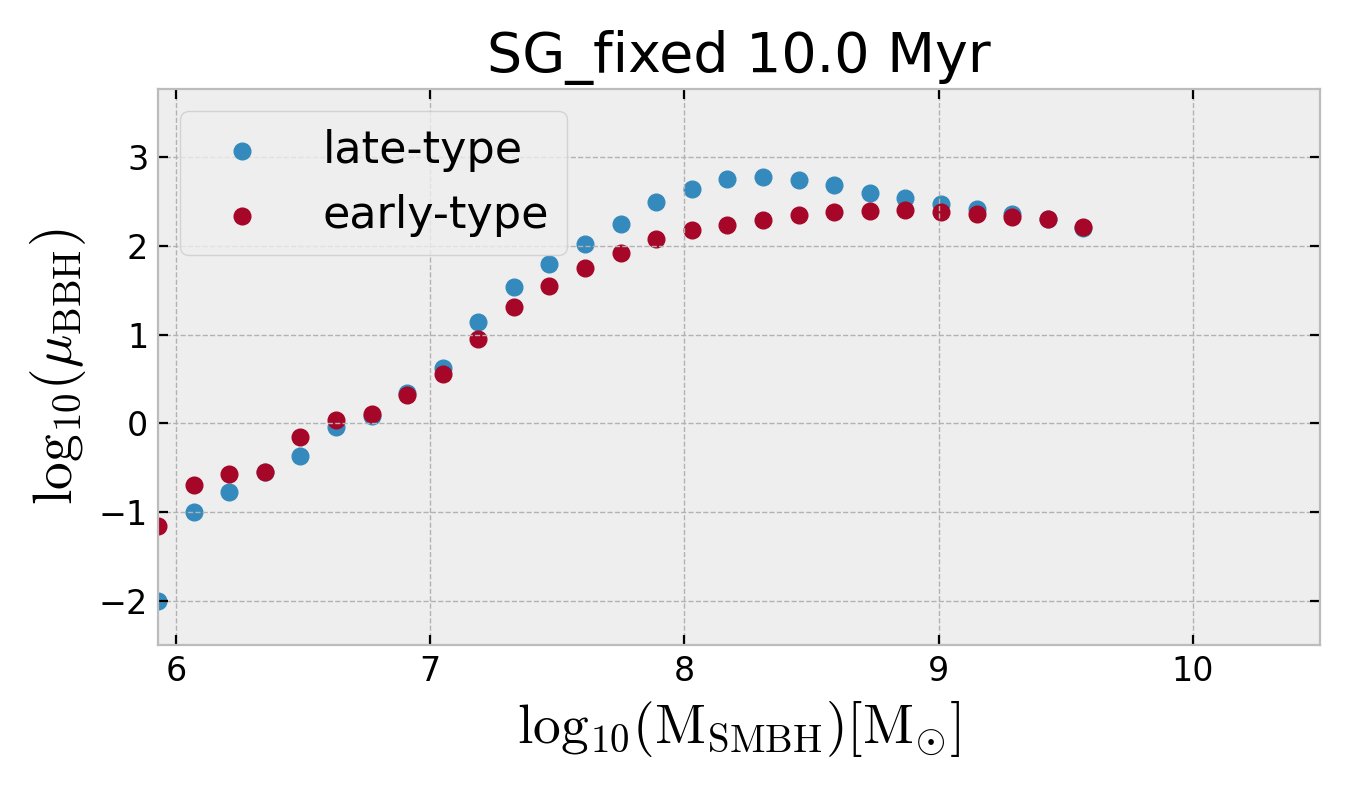}
  \includegraphics[width=3.375 in]{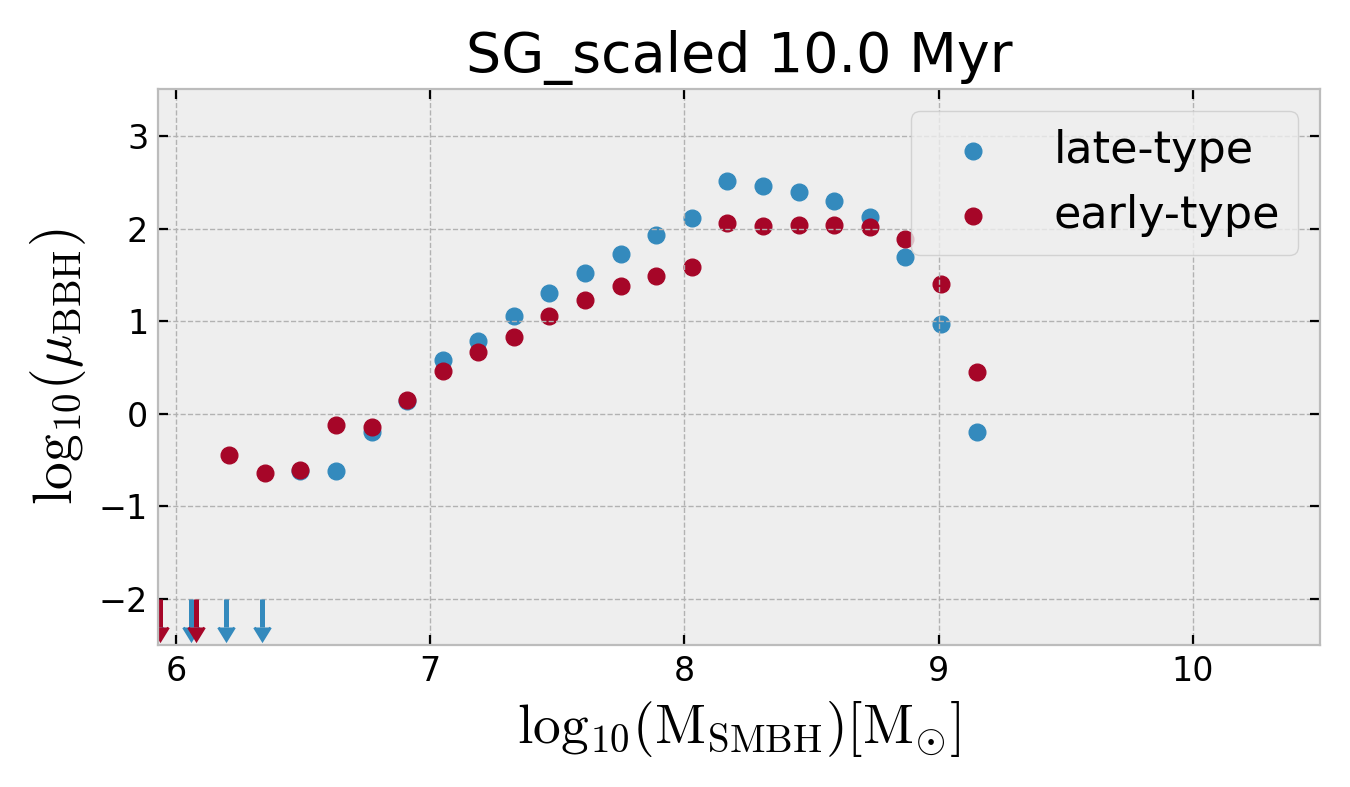}
\caption{\label{fig:bbh-bins}
    \textbf{Expected number of BBH mergers per AGN:}
    Each panel shows the average number of BBH mergers predicted per galaxy
        realization for an AGN episode of a given duration,
        as a function of SMBH mass in early and late-type galaxies.
    100 galaxy realizations are simulated in each of up to 26 mass bins.
    Arrows indicate that a particular mass bin observed zero mergers
        out of 100 realizations.
    Panels on the left are simulations using the default SG disk interpolation
        for \mcfacts{}, with fixed outer disk radius.
    Panels on the right are simulations using SG disk models
        generated using \pagn{}, with the outer disk radius determined
        by an opacity drop.
}
\end{figure*}
Figure \ref{fig:bbh-bins} shows how the expected number $\RateDetection_{\rm BBH}$ of BBH mergers in an AGN disk varies with SMBH mass.  To construct
this figure, we consider  66 different choices
    of specific host galaxy properties for a given disk model.
These choices represent galaxies in 33 stellar mass bins between 
    $10^9$ and $10^{13} \msun$, uniformly distributed in log.
For each galaxy, the SMBH mass is set according to (Eq. \ref{eq:smbh-mass}) for each bin,
    and the NSC mass is set according to (Eq. \ref{eq:nsc-mass})
    for early- and late-type galaxies respectively.
For each mass bin, 100 galaxy realizations are simulated for early-
    and late-type galaxies.  The quantity $\RateDetection_{\rm BBH}$  represents the average number of merging BBH seen in those
    100 galaxy realizations.
Section \ref{sec:disk} describes the different assumptions
    made for various quantities in different mass bins,
    for both the \SGfixed{} and \SGscaled disk model choices.

We find that for \SGfixed{}, \mcfacts{} runs correctly
    for $M_{\mathrm{SMBH}} \leq 10^{9.57} \msun$ (
    corresponding to $M_* \leq 10^{12.25} \msun$).
We find that for \SGscaled, \mcfacts{} runs correctly
    for $M_{\mathrm{SMBH}} \leq 10^{9.15} \msun$
    (corresponding to $M_* \leq 10^{11.875} \msun$).
Appendix \ref{ap:performance} details the performance of \mcfacts{}
    for both disk models.
In practice, as our GSMF finds no sample galaxies with a stellar mass
    greater than $10^{12} \msun$, this does not impact our synthetic universe.

We currently assume a single phase of AGN activity without regard
    to the history of a host galaxy.  We further conservatively assume that after each AGN epoch,  all
    binaries still present in the former AGN disk rapidly  become unbound, e.g. due to ionization interactions with the
    ambient NSC. 
Therefore, by simulating each galaxy realization for a full
    10 Myr, and disregarding BBH mergers after a given time,
    we can describe a BBH merger population for any given lifetime
    without running a new simulation for each AGN lifetime we wish to study.

From Figure \ref{fig:bbh-bins}, we notice
    a preferred SMBH mass scale for both disk-models explored in this publication.
For mergers predicted using the \SGfixed{} disk model,
    this mode appears to be centered between
    $10^8$ and $10^9 \msun$.
As time goes on, we see more BBH mergers further from the preferred
    mass scale.
For shorter AGN lifetimes,
    the shape of $\RateDetection_{\rm BBH}$ versus SMBH mass is rather different.
For long AGN lifetimes, 
    the \SGfixed model produces relatively more merging BBH around higher-mass SMBH,
    compared to shorter AGN lifetimes.
We also note that the overal number of BBH mergers predicted for each galaxy realization
    is much lower for the \SGscaled disk model for shorter AGN lifetimes.
The high rates predicted for the \SGfixed{} disk model
    may be explained by a higher density in the outer disk.

In Figure \ref{fig:bbh-bins}, we notice the most dramatic difference between
    the 0.5 Myr and 1 Myr AGN lifetimes for the \SGfixed{} disk model.
We also notice that the predicted merger rate for the \SGscaled disk model
    increases most after 2 Myr.

\subsection{The intrinsic merging BBH population}
\begin{figure}
\centering
\includegraphics[width=3.3 in]{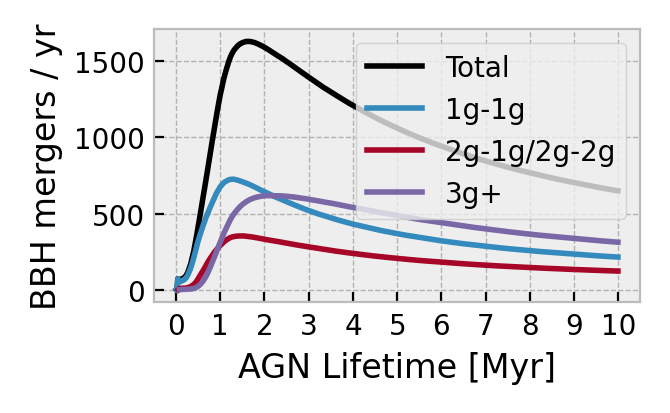}
\includegraphics[width=3.3 in]{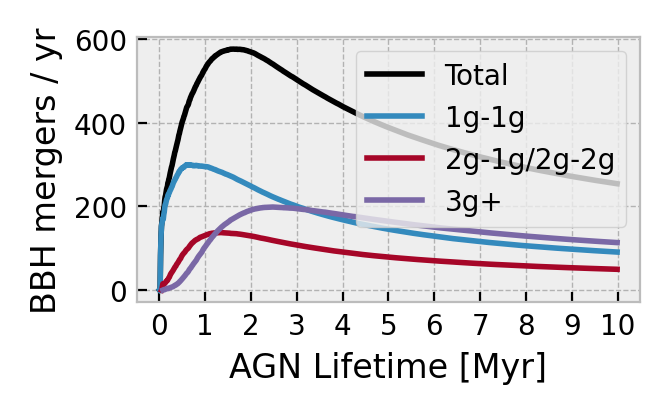}
\caption{\label{fig:rate-intrinsic}
    \textbf{Impact of AGN lifetime on merger rates:}
    The expected number of mergers per year within redshift 2 in a synthetic universe for two disk models,
        as a function of AGN lifetime. 
(top) The \SGfixed{} disk model.
    (bottom) The \SGscaled disk model.
    We see the overall merger rate, as well as first generation, second generation, and higher generation
        merger rates.
}
\end{figure}
\begin{figure}
\centering
\includegraphics[width=3.3 in]{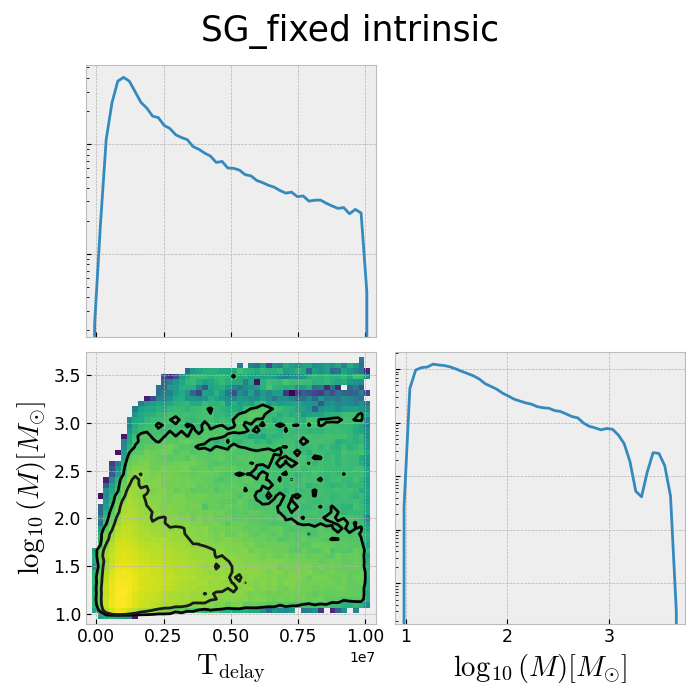}
\includegraphics[width=3.3 in]{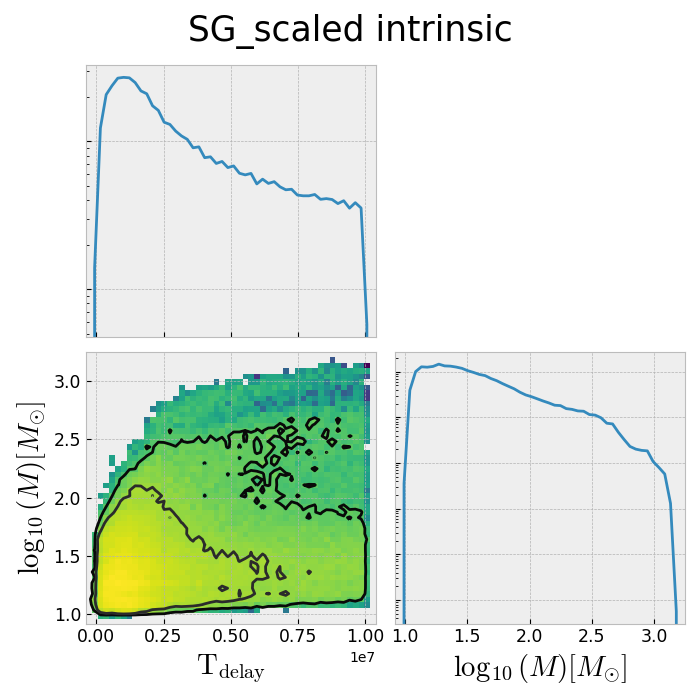}
\caption{\label{fig:delay-time}
    \textbf{Correlation between delay time and mass:}
    A corner plot in delay time and $\log_{10}$ of total mass 
    (of the binary)
        for our two disk models.
    Diagonal plots show one-dimensional histograms for the chosen coordinate,
        while off-diagonal plots show two-dimensional log-scale histograms for chosen coordinates.
    Histograms are weighted by the intrinsic weight for each sample binary (
    Eq. \ref{eq:weight-intrinsic}, and the color scale for the two-dimensional histogram
    reflects these weights in log-scale).
    68\% and 95\% confidence intervals have been identified with contours.
    (top) The \SGfixed{} disk model.
    (bottom) The \SGscaled disk model.
}
\end{figure}

\begin{figure*}
  \centering
  \includegraphics[width=3.375 in]{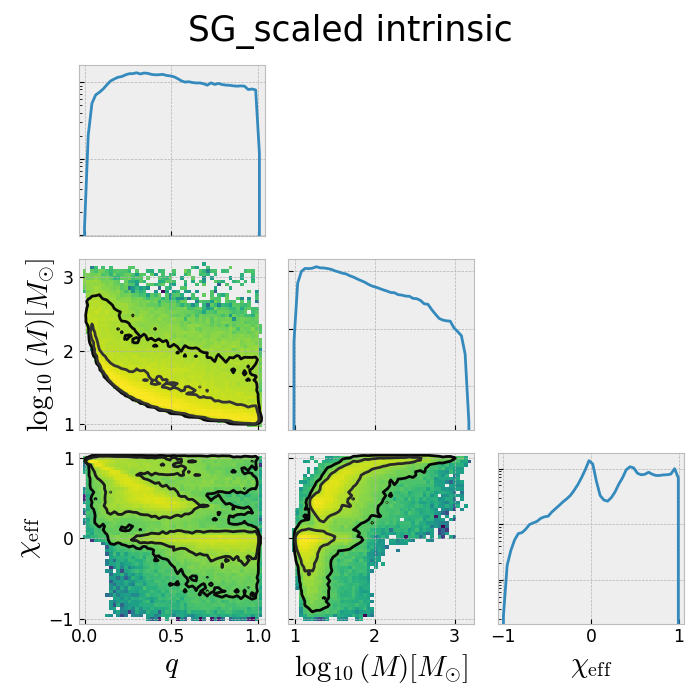}
  \includegraphics[width=3.375 in]{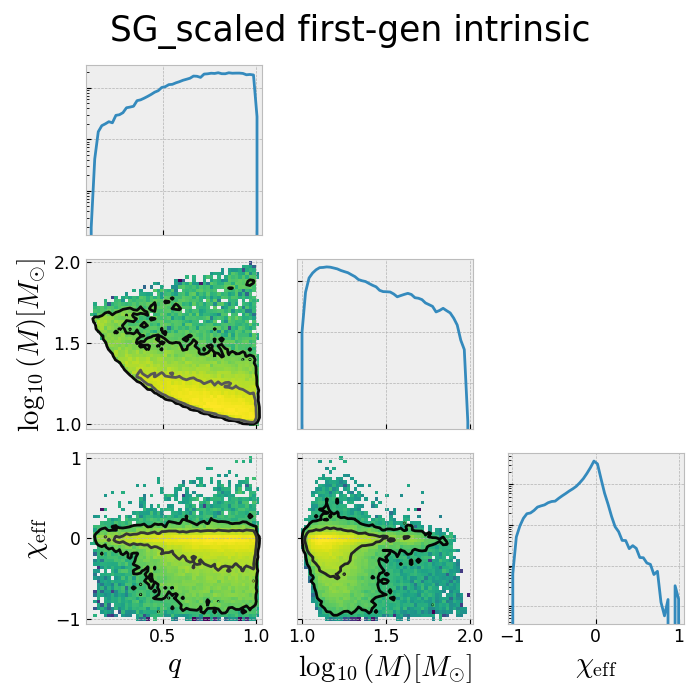}
  \includegraphics[width=3.375 in]{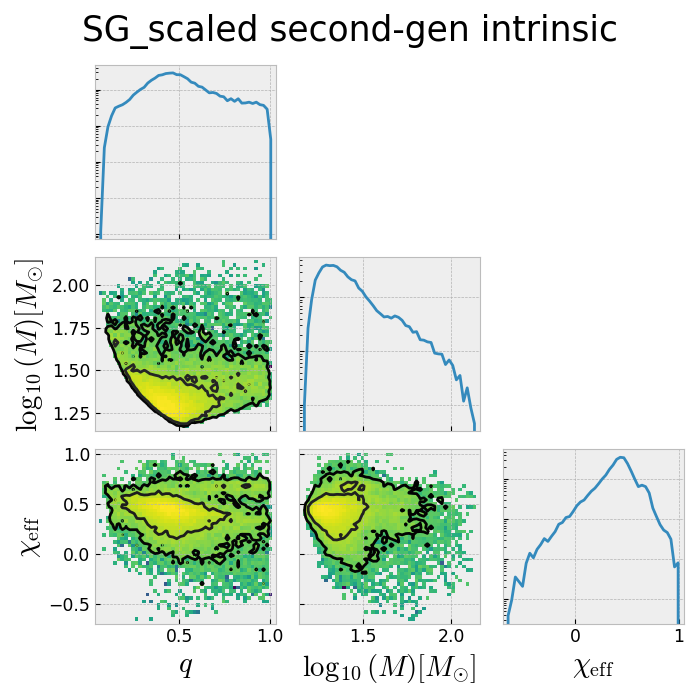}
  \includegraphics[width=3.375 in]{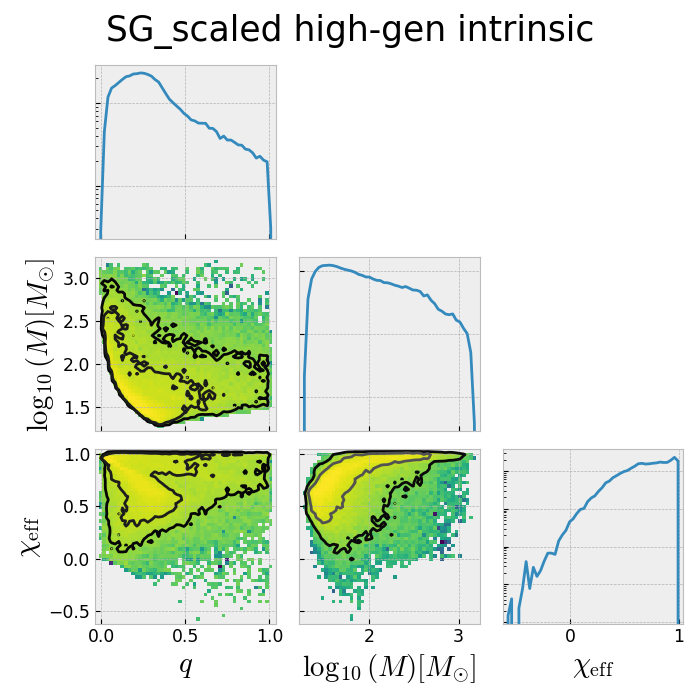}
\caption{\label{fig:corner-pagn-intrinsic}
    \textbf{Properties of BBH mergers in the AGN disk:}
    Each panel shows a corner plot in mass ratio,
        $\log_{10}$ of total mass, and $\chi_{\mathrm{eff}}$.
    Diagonal plots show one-dimensional histograms for the chosen coordinate,
        while off-diagonal plots show two-dimensional log-scale histograms for chosen coordinates.
    Histograms are weighted by the intrinsic weight for each sample binary (
    Eq. \ref{eq:weight-intrinsic}, and the color scale for the two-dimensional histogram
    reflects these weights in log-scale).
    68\% and 95\% confidence intervals have been identified with contours.
    All four images visualize BBH mergers predicted in the first 10 Myr of AGN lifetime
        for the \SGscaled disk model.
    (top-left): All mergers.
    (top-right): 1g-1g mergers.
    (bottom-left): 2g-1g and 2g-2g mergers.
    (bottom-right): higher generation mergers.
}
\end{figure*}

Following Section \ref{sec:postproc},
    we estimate the intrinsic AGN population in a synthetic universe
    by sampling the properties of a galactic population (see Figure \ref{fig:orange}),
    categorize galaxy samples as early- or late-type,
    and normalize according to an assumed AGN density (see Section \ref{sec:postproc}).
Figure \ref{fig:rate-intrinsic} shows the expected BBH merger rate for binaries
    in a synthetic universe as a function of AGN lifetime
    for our two disk models
    (see Section \ref{sec:disk} for the details of each disk model).
Two competing effects are at work as AGN lifetime increases:
    mergers which form only after a longer delay since AGN activity commences (henceforth denoted the delay time) are allowed to contribute to the population,
    and yet the contribution of individual mergers is reduced
    as our duty cycle increases (see Eq. \ref{eq:duty-cycle}).
While these intrinsic merger rates are not yet comparable with
    LVK detection rates (see Section \ref{sec:results-detection}
    for such an analysis), they immediately highlight several important insights:
    i) there is a substantial population of merging BBHs predicted from AGN
        in a realistic population of galaxies;
    ii) there is a peak in merger events within the first two Myr
        of an AGN cycle;
    iii) different assumptions about the nature of AGN disks can change 
        the simulated merger rate by at least an order of magnitude.

Figure \ref{fig:delay-time} illustrates the correlation of delay time and
    total mass in the intrinsic population of BBH mergers
    predicted for our synthetic universe with our two disk models.
In each case, the highest mass binaries begin to merge after the peak
    of merger activity for a given AGN cycle.
Here again, we highlight the impact of the AGN disk model on the predicted
    population of BBH mergers.
For our \SGfixed{} disk model,
    the most massive BBH mergers have a higher mass than for the
    \SGscaled{} disk model.

To reiterate the impact of AGN duty cycle,
    recall that the number density of AGN
    which are actively accreting (see Section \ref{sec:postproc}) is fixed,
    and therefore different AGN lifetime assumptions also
    require different duty-cycle assumptions.
In other words, if our AGN lifetime assumption halves,
    we assume there are twice as many AGN cycles in our
    100 Myr epoch.
Comparing Figure \ref{fig:rate-intrinsic} and the one-dimensional
    histogram in delay time for Figure \ref{fig:delay-time}
    reveals the impact of our duty cycle assumption.
When there is a very strong peak in delay time,
    the intrinsic merger rate falls off steeply after that peak.
Conversely, the intrinsic merger rate can depend only loosely on
    AGN lifetime when
    the peak is less dominant.

Figure \ref{fig:corner-pagn-intrinsic} illustrates the properties of merging binaries
    in the \SGscaled disk model.
Different generational modes (1g-1g, 2g-1g, 2g-2g, 3g+) can be visually identified in
    the overall population.
Characteristic of many dynamic formation channels,
    we find a preference for equal mass BBHs.
We note, however, that this preference is not as strong as for isolated binary evolution
    (see Figure \ref{fig:bbh-nal}).
Consistent with \citet{mcfactsI}, we notice a small bump in the total mass histogram
    near $70 \msun$, which is unsurprising considering our IMF.
It is likewise unsurprising that higher generational mergers tend toward higher mass.

Notably, this figure demonstrates that merging BBH with masses $\simeq 10^3 \msun$ 
    can be produced in AGN disks (top left panel),
    primarily through high-generation mergers (bottom right).
Of course, as seen in Figure \ref{fig:delay-time},
    the largest masses require long AGN lifetimes,
    building up only slowly within the \SGscaled model.
Above about $70\msun$, the mass distribution is roughly a power law.
Though the largest BBH only occur after long delay times,
    even for lifetimes as short as $\simeq 2$ Myr our AGN disks produce mergers of
    $\simeq 100 \msun$ black holes.  

Consistent with \citet{mcfactsII},
    we identify modes of higher spin and an anti-correlation between $q$ and $\chi_{\rm eff}$
    for higher generation mergers in Figure \ref{fig:corner-pagn-intrinsic}.
First, we identify a mode along $\chi_{\rm eff} = 0$ in the first generation
    population due to the random spin initialization.
BHs on eccentric orbit acquire negative spin during circularization,
    causing many first generation mergers to have $\chi_{\rm eff} < 0$.
Some values reach as low as $\chieff = -1$,
    indicating both BHs' spins are anti-aligned to the binary angular momentum.
First generation mergers encounter a lower limit of $q \sim 0.15$
    due to the nature of our BH IMF.
The effective spin of some second generation mergers
    between similar component masses ($q \sim 1$) remains negative,
    but the majority are now positive.
Negative values are far less likely for low $q$ mergers.
As the system evolves,
    the more massive component is increasingly likely to be a merger remnant
    and thus have a spin aligned with the disk. This is because the spin angular momentum of the remnant is preferentially aligned
    with the orbital angular momentum of the progenitor binary systems
    \cite{Tichy2008}.

High generation mergers further express the trend of spin alignment,
    as vey few events occur with negative effective spin.
Mergers of any mass ratio can be almost completely aligned with the disk
    ($\chi_{\rm eff} \sim 1$).

\subsection{Comparison of simulated population with LVK observations}
\label{sec:results-detection}

\begin{figure}
\centering
\includegraphics[width=3.3 in]{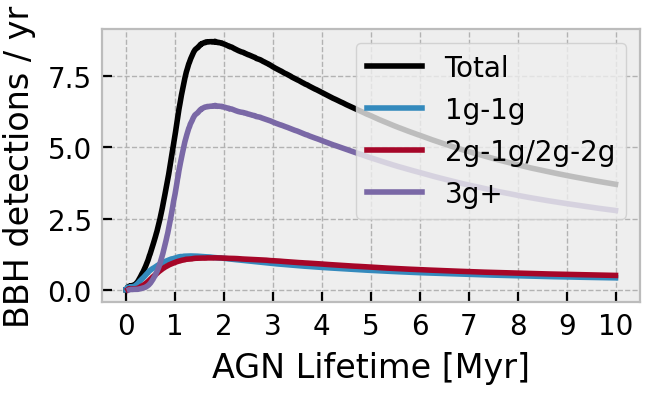}
\includegraphics[width=3.3 in]{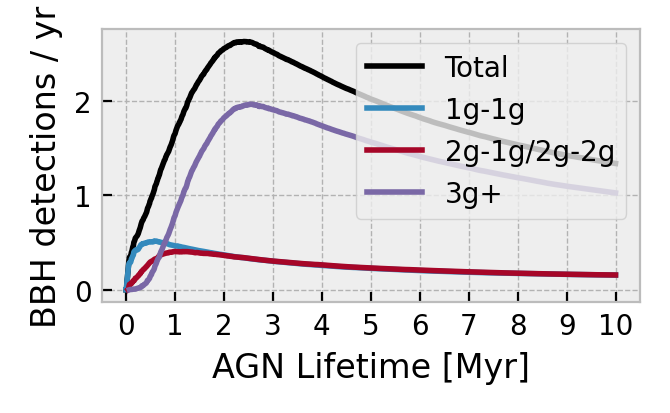}
\caption{\label{fig:rate-obs}
    \textbf{Impact of AGN lifetime on LVK detection rates:}
    The expected number of LVK detections for each disk model
        as a function of AGN lifetime
        (assuming the sensitivity of the third observing run for ground-based 
        gravitational wave instruments).
    This observable merger rate represents a volume out to redshift 2.
    (top) The \SGfixed{} disk model.
    (bottom) The \SGscaled disk model.
}
\end{figure}

\begin{figure}
\centering
\includegraphics[width=3.3 in]{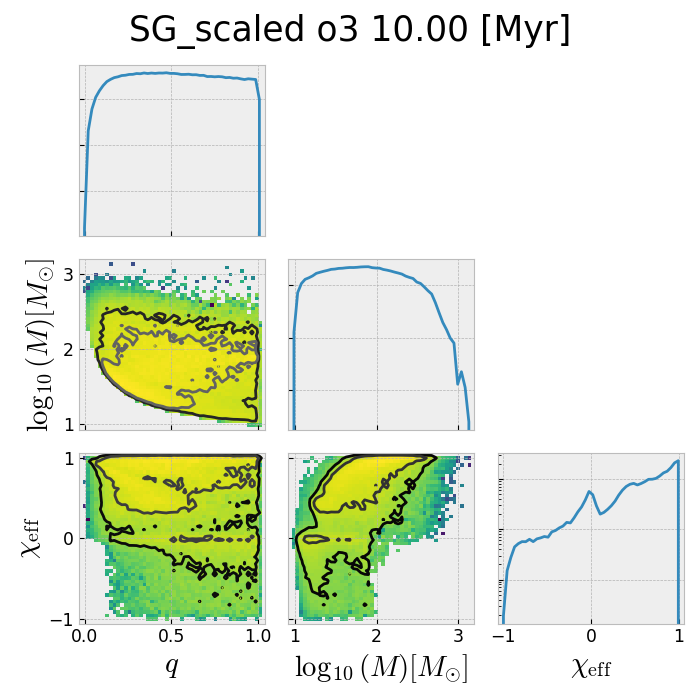}
\caption{\label{fig:corner-pagn-detection}
    \textbf{Observable population of BBH mergers:}
    Compare with Figure \ref{fig:corner-pagn-intrinsic}.
Here, we weigh each merger sample by its
LVK detection rate  (i.e. Eq. \ref{eq:detection-rate}), rather than
intrinsic merger rate  (i.e. Eq. \ref{eq:weight-intrinsic}).
This figure represents mergers in a $10$ Myr AGN lifetime.
}
\end{figure}

Figure \ref{fig:rate-obs} estimates LVK detection rates for 
    a sensitivity matching the third observing run of ground-based instruments
    (following Section \ref{sec:detection}).
The figure explores both of our two disk models as a function of AGN lifetime.
In order to explain why predicted rates may decrease as a function of AGN lifetime,
    recall that for a fixed, observationally constrained, AGN number density,
    a higher AGN lifetime implies a larger duty cycle (and therefore fewer new episodes of AGN activity;
    See Eq. \ref{eq:duty-cycle}).
We can immediately conclude that our assumed disk model
    and AGN lifetime influences LVK detection rates predicted from 
    \mcfacts{} by at least one order of magnitude.
Considering only the total event count predicted versus observed (roughly 90),
    at face value, we expect between 0 and 10 percent of LVK observations
    may occur in an AGN disk.
We would like to add a caveat in that this may be best interpreted as a lower limit,
    as in this work we do not account for dynamic captures of external stellar objects
    being captured by and added the disk over time.

Figure \ref{fig:corner-pagn-detection} illustrates the properties of the
    observable population in our fiducial model (\SGscaled).
Interestingly, no features of the overall population 
    disappear, comparing to  Figure \ref{fig:corner-pagn-intrinsic}.
Conversely, higher mass mergers which occur less frequently are weighted higher
    by our detection model.
Therefore, we predict a more even spread of parameters than
    for the intrinsic population.
We conclude that sparse features of the intrinsic population
    may be more important after the application of a detection model.

\begin{figure*}
\centering
\includegraphics[width=3.3 in]{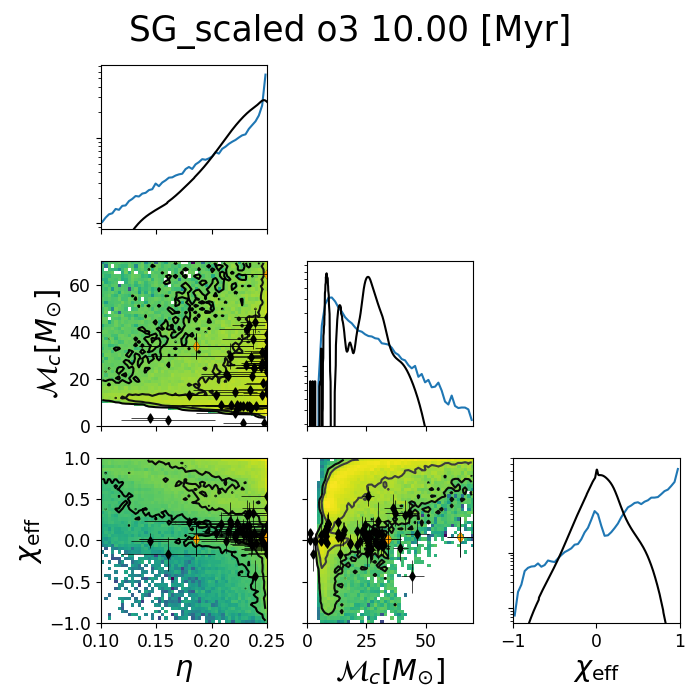}
\includegraphics[width=3.3 in]{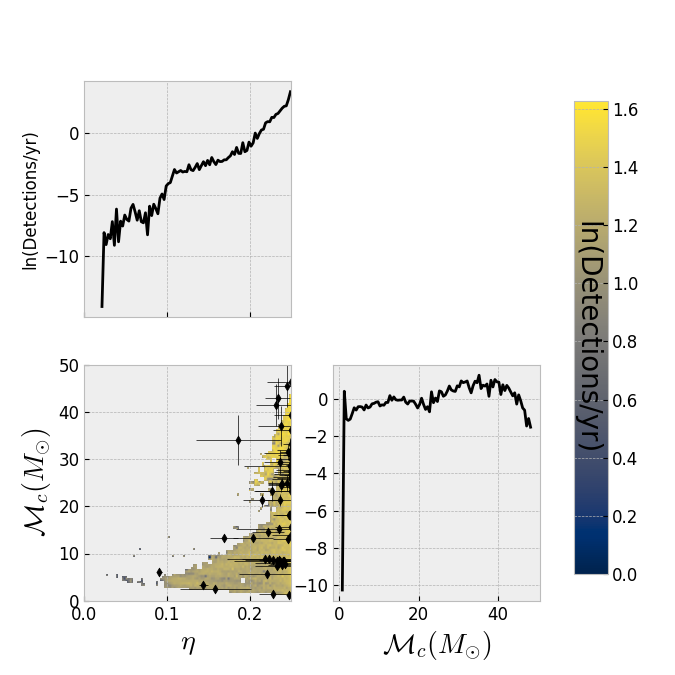}
\caption{\label{fig:bbh-nal}
    \textbf{Comparison with confident events in the GWTC-3:}
    (left) A corner plot 
        (similar to Figure \ref{fig:corner-pagn-detection})
        in symmetric mass ratio, source-frame
        chirp mass, and effective spin
        for the predicted AGN disk BBH detections with the \SGscaled disk model.
        Histograms are weighted by detectability (see Eq. \ref{eq:detection-rate}).
        Here, we have zoomed in on the sub-population where
            $\eta \in [0.1,0.25]$ and $\mathcal{M}_c \in [0,70] \msun$.
    (right) A similar corner plot in symmetric mass ratio and source-frame
        chirp mass for an isolated binary population
        presented in \citet{st_inference_interp}.
    Scattered on top of the two-dimensional histograms
        are the maximum likelihood estimate and one-sigma
        error for a truncated Gaussian fitting each
        of the confident LVK observations in the GWTC-3 catalog.
    Black lines in the diagonals of the left corner plot
        visualize a one-dimensional mixture of Gaussian components
        for the confident events in the GWTC-3 catalog.
    GW190521 and GW190929\_012149 are orange, to highlight their significance.
}
\end{figure*}

Figure \ref{fig:bbh-nal} shows the merger population for the \SGscaled disk model
    in new coordinates:
    symmetric mass ratio ($\eta = (m_1m_2)/(m_1 + m_2)^2$)
    and source-frame chirp mass ($\mathcal{M}_c = \eta^{3/5}(m_1 + m_2)$).
These coordinates are better constrained by the waveform,
    and therefore are more natural for directly comparing observed data to
    simulated populations.
This figure features the same merger population as Figure \ref{fig:corner-pagn-detection},
    zoomed in to $\eta \in [0.1,0.25]$ and $\mathcal{M}_c \in [5,70] \msun$,
    to highlight the agreement between the predicted merger observations
    and the confident LVK observations in the first three observing runs
    \citep[consistent with][]{LIGO-O3-O3b-RP}.
A maximum likelihood and standard deviation for each of these confident detections
    is estimated using a truncated Gaussian fit to parameter estimation samples
    \citep{nal-methods-paper}.

From Figure \ref{fig:bbh-nal}, we notice good qualitative agreement between our model and confident events
    above $20 \msun$, even away from equal mass.
Our model also suggests detections would be expected in regions currently lacking confident detections, such as binaries
with very large effective spin.
Our black hole IMF begins at $5 \msun$, and therefore cannot explain observations
    below $\mathcal{M}_c = 4.325 \msun$
    (such as many which possibly contain neutron stars).
This IMF is informed by LVK observations,
    and therefore may introduce a significant source of bias.
Future work will explore alternative black hole IMF assumptions.

For reference, the right side of Figure \ref{fig:bbh-nal} shows
    a similar comparison, between O3 observations and 
    a model based on isolated binary stellar evolution.
This model was the preferred model of \citet{st_inference_interp},
    incorporating the delayed supernova engine \citep{Fryer2012},
    weak pair-instability effects, a 5 \% Bondi-Hoyle accretion rate,
    a $2.5 \msun$ maximum neutron star mass, 
    a common envelope ejection efficiency of $\alpha_{\mathrm{CE}} = 1.0$, 
    efficient dynamically stable Roche-lobe overflow ($f_a = 0.922$),
    a supernova recoil kick from a single Maxwellian described by $\sigma_{\mathrm{kick}} = 108.3 \mathrm{km/s}$
    (not reduced by fallback),
    and reduced mass loss from stellar wind during the hydrogen-dominated parts of a star's life
    (a factor of $f_{\mathrm{WIND1}} = 0.328$ compared to the standard \citet{Vink2001} power law).
    To read more about these assumptions, refer to \citet{st_inference_interp}.
    
As seen in this figure,
    this isolated binary evolution model is consistent with most GW observations, though in modest
    tension with the high-mass population reported to date.

Qualitatively, Figure \ref{fig:bbh-nal} may lend insight to the preferred
    formation channel for particular events.
Specifically, the AGN channel may not be likely to produce BBH mergers below
    a source-frame chirp mass of
    $\mathcal{M}_c = 4.325$ $\msun$.
In addition, a few events which stand apart from the isolated binary population
    channel may originate from the AGN disk channel.
The cosmic origin of massive compact binary mergers such as GW190521
    have long been debated
    \citep{Gayathri2022, Gamba2023Nature, MortonGW190521, LIGO-O3-O3a-RP, LIGO-O3-O3b-RP,
    RomeroShawGW190521, Fragione_2020, Belczynski2020-EvolutionaryRoads}.
In this work, we predict that the AGN channel can produce events as massive as
    this event,
    although this particular event appears at the fringes of the distribution.
Additionally, GW190929\_012149 appears to be more consistent with the AGN
    population predicted here than with the isolated binary channel.
We reserve a full comparison of BBH merger populations from AGN disk
    and isolated binary environments for future work.
 \section{Discussion}
\label{sec:discussion}
In this work, we present simulated BBH merger populations from
    AGN disk environments representing 132 different disk models:
    33 different SMBH mass assumptions for early- and late-type galaxies,
    for each of two disk models which scale according to SMBH mass.
We have sampled from the properties of galaxies in different periods
    of the history of the Universe, in order to populate 
    a synthetic universe with a realistic population of AGN.
Further, we have estimated the population of double compact object
    mergers from the AGN disk formation channel,
    which modern ground-based gravitational wave observatories might detect.

In short, 
    we have produced the first work modeling the AGN channel for 
    BBH mergers in a universe with galaxy properties similar to our own, 
    and done so using a realistic set of models 
    for both the AGN disk and NSC appropriate to each galaxy. 
Such a comprehensive cosmological perspective has enabled us to 
    make several critically interesting and new inferences:
\begin{itemize}
\item Merging BBH populations likely have a preferred SMBH 
    mass scale near $10^8 \msun$,
    due to the fact that such SMBH are sufficiently numerous in the universe
    \textit{and} have large enough NSCs to create many merging BBH.
\item The opacity of AGN disks can substantially impact predicted merger rates 
    (by an order of magnitude) 
    through the influence of opacity on migration torques 
    and the effects of feedback on migrating objects.
    Importantly, higher merger rates occur in higher opacity disks.
\item The BBH merger rate in an AGN disk depends strongly on the
    lifetime of the the AGN disk,
    with a significant peak in the delay time of individual mergers around 1-2 Myr.
    AGN lifetimes shorter than 1 Myr may not contribute substantially to
    the LVK detections.
    AGN lifetimes longer than 2 Myr may predict fewer BBH mergers in
    a universe like our own.
\item Populations of first, second, and higher generation BBH mergers in 
    an AGN disk may have quite disparate properties and 
    appear as separable populations -- not only in mass, but also mass ratio, spin.
\item Filtering the population of BBH mergers in an AGN disk by detectability in the LVK's third observing run
    does not appear to miss distinct sub-populations
    (i.e. Figure \ref{fig:bbh-nal} does not appear to be
    missing any modes of the distribution in the top-left panel of Figure \ref{fig:corner-pagn-intrinsic}).
\item BBH mergers in an AGN disk can account for LVK observations such as 
    GW190521 and GW190929\_012149. In its present form, however, \mcfacts{} was not able to easily reproduce
        the high mass and modest effective spin of GW190521;
        however, as operating points change and particularly as new physics
        is added to the \mcfacts{} simulation framework,
        the parameter space which \mcfacts{} outputs will evolve alongside the code.
\item This study supports past work in affirming
    that AGN disks are a formation channel which can
    account for the most massive compact binary mergers reported by the LVK in
    GWTC-3 \citep{GWTC-3}.
\end{itemize}

The mass and spin predictions from these models
    overlap with many of the LVK's previous detections.
In principle, many of them could arrive from this channel.
\citet{mcfactsII} showed the \mcfacts{}
    predictions for spin in merging BBH binaries for individual galaxies.
Alongside that, by studying a population drawn from AGN representing
    an evolving galactic stellar mass function,
    we effectively marginalize that population over AGN
    in a universe like our own.

This work was motivated by a previous paper \citep{st_inference_interp},
    which compared models of isolated stellar evolution to observations.
In the future, we can use the same hierarchical inference methods
    to refine AGN formation models for double compact objects,
    and even designate probable formation channels for specific LVK observations.
We must always be careful not to over-interpret population synthesis studies
    \citep{COSMICZevin2021,Cheng2023},
    but with attention to detail we can use these tools to better understand
    the impact of our assumptions on the population of compact binaries
    and work toward a more complete understanding of their formation channels.
 
\begin{paragraph}{Data/Code Availability}
The data used in this paper is available at \url{https://zenodo.org/records/13993722}
\citep{mcfactsIIIdata}.
\end{paragraph}

\begin{acknowledgments}
VD is supported by an appointment to the NASA Postdoctoral Program at the NASA Goddard Space Flight Center administered by Oak Ridge Associated Universities under contract NPP-GSFC-NOV21-0031.
BM, KESF and HEC are supported by NSF AST-2206096. BM \& KESF are supported by NSF AST-1831415 and Simons Foundation Grant 533845 as well as Simons Foundation sabbatical support and support to release this code. The Flatiron Institute is supported by the Simons Foundation.
KN acknowledges support from the LSSTC Data Science Fellowship Program, which is funded by LSSTC, NSF Cybertraining Grant no. 1829740, the Brinson Foundation, and the Moore Foundation.
ROS acknowledges support from NSF PHY 2012057, PHY 2309172, and AST 2206321.

We acknowledge software packages used in this publication,
    including
    \texttt{NumPy} \citep{harris2020array},
    \texttt{ScirPy} \citep{2020SciPy-NMeth}, 
    \texttt{Matplotlib} \citep{Hunter_2007}, 
\texttt{Astropy} \citep{astropy:2013,astropy:2018,astropy:2022}, 
    \texttt{lalsuite} \cite{lalsuite},
    and \texttt{H5py} \citep{collette_python_hdf5_2014}.
This research was done using resources provided by the 
    Open Science Grid \citeOSG,
    which is supported by the National
    Science Foundation awards \#2030508 and \#1836650,
    and the U.S. Department of Energy's Office of Science. 
\end{acknowledgments}

\bibliography{Bibliography.bib}

\begin{thebibliography}{}
\expandafter\ifx\csname natexlab\endcsname\relax\def\natexlab#1{#1}\fi
\providecommand{\url}[1]{\href{#1}{#1}}
\providecommand{\dodoi}[1]{doi:~\href{http://doi.org/#1}{\nolinkurl{#1}}}
\providecommand{\doeprint}[1]{\href{http://ascl.net/#1}{\nolinkurl{http://ascl.net/#1}}}
\providecommand{\doarXiv}[1]{\href{https://arxiv.org/abs/#1}{\nolinkurl{https://arxiv.org/abs/#1}}}

\bibitem[{Abbott {et~al.}(2019)Abbott, Abbott, Abbott, Abraham, \&
  et~al.}]{GWTC-1}
Abbott, B., Abbott, R., Abbott, T., Abraham, S., \& et~al. 2019, Physical
  Review X, 9, 031040

\bibitem[{Abbott {et~al.}(2021)Abbott, Abbott, Abraham, Acernese, \&
  et~al.}]{GWTC-2}
Abbott, R., Abbott, T., Abraham, S., Acernese, F., \& et~al. 2021, Physical
  Review X, 11, 021053

\bibitem[{Ade {et~al.}(2016)}]{Planck2015}
Ade, P. A.~R., {et~al.} 2016, Astron. Astrophys., 594, A13,
  \dodoi{10.1051/0004-6361/201525830}

\bibitem[{{Arca Sedda} {et~al.}(2023){Arca Sedda}, {Naoz}, \&
  {Kocsis}}]{ArcaSedda23}
{Arca Sedda}, M., {Naoz}, S., \& {Kocsis}, B. 2023, Universe, 9, 138,
  \dodoi{10.3390/universe9030138}

\bibitem[{{Astropy Collaboration} {et~al.}(2018){Astropy Collaboration},
  {Price-Whelan}, {Sip{\H{o}}cz}, {G{\"u}nther}, \& et~al.}]{astropy:2018}
{Astropy Collaboration}, {Price-Whelan}, A.~M., {Sip{\H{o}}cz}, B.~M.,
  {G{\"u}nther}, H.~M., \& et~al. 2018, \aj, 156, 123,
  \dodoi{10.3847/1538-3881/aabc4f}

\bibitem[{{Astropy Collaboration} {et~al.}(2013){Astropy Collaboration},
  {Robitaille}, {Tollerud}, {Greenfield}, \& et~al.}]{astropy:2013}
{Astropy Collaboration}, {Robitaille}, T.~P., {Tollerud}, E.~J., {Greenfield},
  P., \& et~al. 2013, \aap, 558, A33, \dodoi{10.1051/0004-6361/201322068}

\bibitem[{{Astropy Collaboration} {et~al.}(2022){Astropy Collaboration},
  {Price-Whelan}, {Lim}, {Earl}, {Starkman}, {Bradley}, {Shupe}, {Patil},
  {Corrales}, {Brasseur}, {N{\"o}the}, {Donath}, {Tollerud}, {Morris},
  {Ginsburg}, {Vaher}, {Weaver}, {Tocknell}, {Jamieson}, {van Kerkwijk},
  {Robitaille}, {Merry}, {Bachetti}, {G{\"u}nther}, {Aldcroft},
  {Alvarado-Montes}, {Archibald}, {B{\'o}di}, {Bapat}, {Barentsen},
  {Baz{\'a}n}, {Biswas}, {Boquien}, {Burke}, {Cara}, {Cara}, {Conroy},
  {Conseil}, {Craig}, {Cross}, {Cruz}, {D'Eugenio}, {Dencheva}, {Devillepoix},
  {Dietrich}, {Eigenbrot}, {Erben}, {Ferreira}, {Foreman-Mackey}, {Fox},
  {Freij}, {Garg}, {Geda}, {Glattly}, {Gondhalekar}, {Gordon}, {Grant},
  {Greenfield}, {Groener}, {Guest}, {Gurovich}, {Handberg}, {Hart},
  {Hatfield-Dodds}, {Homeier}, {Hosseinzadeh}, {Jenness}, {Jones}, {Joseph},
  {Kalmbach}, {Karamehmetoglu}, {Ka{\l}uszy{\'n}ski}, {Kelley}, {Kern},
  {Kerzendorf}, {Koch}, {Kulumani}, {Lee}, {Ly}, {Ma}, {MacBride}, {Maljaars},
  {Muna}, {Murphy}, {Norman}, {O'Steen}, {Oman}, {Pacifici}, {Pascual},
  {Pascual-Granado}, {Patil}, {Perren}, {Pickering}, {Rastogi}, {Roulston},
  {Ryan}, {Rykoff}, {Sabater}, {Sakurikar}, {Salgado}, {Sanghi}, {Saunders},
  {Savchenko}, {Schwardt}, {Seifert-Eckert}, {Shih}, {Jain}, {Shukla}, {Sick},
  {Simpson}, {Singanamalla}, {Singer}, {Singhal}, {Sinha}, {Sip{\H{o}}cz},
  {Spitler}, {Stansby}, {Streicher}, {{\v{S}}umak}, {Swinbank}, {Taranu},
  {Tewary}, {Tremblay}, {de Val-Borro}, {Van Kooten}, {Vasovi{\'c}}, {Verma},
  {de Miranda Cardoso}, {Williams}, {Wilson}, {Winkel}, {Wood-Vasey}, {Xue},
  {Yoachim}, {Zhang}, {Zonca}, \& {Astropy Project
  Contributors}}]{astropy:2022}
{Astropy Collaboration}, {Price-Whelan}, A.~M., {Lim}, P.~L., {et~al.} 2022,
  \apj, 935, 167, \dodoi{10.3847/1538-4357/ac7c74}

\bibitem[{{Bartos} {et~al.}(2017){Bartos}, {Kocsis}, {Haiman}, \&
  {M{\'a}rka}}]{Bartos17}
{Bartos}, I., {Kocsis}, B., {Haiman}, Z., \& {M{\'a}rka}, S. 2017, \apj, 835,
  165, \dodoi{10.3847/1538-4357/835/2/165}

\bibitem[{{Bavera} {et~al.}(2020){Bavera}, {Fragos}, {Zevin}, {Berry},
  {Marchant}, {Andrews}, {Coughlin}, {Dotter}, {Kovlakas}, {Misra},
  {Serra-Perez}, {Qin}, {Rocha}, {Rom{\'a}n-Garza}, {Tran}, \&
  {Zapartas}}]{2020arXiv201016333B}
{Bavera}, S.~S., {Fragos}, T., {Zevin}, M., {et~al.} 2020, arXiv e-prints,
  arXiv:2010.16333.
\newblock \doarXiv{2010.16333}

\bibitem[{Belczynski {et~al.}(2020)Belczynski, Klencki, Fields, Olejak, \&
  et~al.}]{Belczynski2020-EvolutionaryRoads}
Belczynski, K., Klencki, J., Fields, C.~E., Olejak, A., \& et~al. 2020,
  Astronomy \& Astrophysics, 636, A104, \dodoi{10.1051/0004-6361/201936528}

\bibitem[{{Bellovary} {et~al.}(2016){Bellovary}, {Mac Low}, {McKernan}, \&
  {Ford}}]{Bellovary16}
{Bellovary}, J.~M., {Mac Low}, M.-M., {McKernan}, B., \& {Ford}, K.~E.~S. 2016,
  \apjl, 819, L17, \dodoi{10.3847/2041-8205/819/2/L17}

\bibitem[{Broekgaarden \& Berger(2021)}]{broekgaarden2021formation}
Broekgaarden, F.~S., \& Berger, E. 2021, The Astrophysical Journal Letters,
  920, L13

\bibitem[{Broekgaarden {et~al.}(2021)Broekgaarden, Berger, Stevenson, Justham,
  Mandel, \& Chru{\'s}li{\'n}ska}]{broekgaarden2021impact}
Broekgaarden, F.~S., Berger, E., Stevenson, S., {et~al.} 2021, arXiv preprint
  arXiv:2112.05763

\bibitem[{{Calcino} {et~al.}(2024){Calcino}, {Dempsey}, {Dittmann}, \&
  {Li}}]{Calcino24}
{Calcino}, J., {Dempsey}, A.~M., {Dittmann}, A.~J., \& {Li}, H. 2024, \apj,
  970, 107, \dodoi{10.3847/1538-4357/ad4a53}

\bibitem[{{Cheng} {et~al.}(2023){Cheng}, {Zevin}, \& {Vitale}}]{Cheng2023}
{Cheng}, A.~Q., {Zevin}, M., \& {Vitale}, S. 2023, \apj, 955, 127,
  \dodoi{10.3847/1538-4357/aced98}

\bibitem[{Collette(2013)}]{collette_python_hdf5_2014}
Collette, A. 2013, Python and HDF5 (O'Reilly)

\bibitem[{Cook {et~al.}(2024)Cook, McKernan, Ford, Delfavero, Nathaniel,
  Postiglione, Ray, \& {O'Shaughnessy}}]{mcfactsII}
Cook, H.~E., McKernan, B., Ford, S., {et~al.} 2024

\bibitem[{Delfavero {et~al.}(2024)Delfavero, Ford, McKernan, Cook, Nathaniel,
  Postiglione, Ray, \& O'Shaughnessy}]{mcfactsIIIdata}
Delfavero, V., Ford, K.~S., McKernan, B., {et~al.} 2024, McFACTS BBH Mergers in
  a Synthetic Universe,  Zenodo, \dodoi{10.5281/zenodo.13993722}

\bibitem[{Delfavero {et~al.}(2023)Delfavero, O'Shaughnessy, Belczynski, Drozda,
  \& Wysocki}]{st_inference_interp}
Delfavero, V., O'Shaughnessy, R., Belczynski, K., Drozda, P., \& Wysocki, D.
  2023, Phys. Rev. D, 108, 043023, \dodoi{10.1103/PhysRevD.108.043023}

\bibitem[{Delfavero {et~al.}(2022)Delfavero, O'Shaughnessy, Wysocki, \&
  Yelikar}]{nal-methods-paper}
Delfavero, V., O'Shaughnessy, R., Wysocki, D., \& Yelikar, A. 2022, Compressed
  Parametric and Non-Parametric Approximations to the Gravitational Wave
  Likelihood,  arXiv, \dodoi{10.48550/ARXIV.2205.14154}

\bibitem[{Di~Carlo {et~al.}(2020)}]{DiCarlo_2020lfa}
Di~Carlo, U.~N., {et~al.} 2020, Mon. Not. Roy. Astron. Soc., 498, 495,
  \dodoi{10.1093/mnras/staa2286}

\bibitem[{Dominik {et~al.}(2015)Dominik, Berti, O’Shaughnessy, Mandel,
  Belczynski, Fryer, Holz, Bulik, \& Pannarale}]{DominikIII}
Dominik, M., Berti, E., O’Shaughnessy, R., {et~al.} 2015, The Astrophysical
  Journal, 806, 263

\bibitem[{Fontana {et~al.}(2006)Fontana, Salimbeni, Grazian, Giallongo,
  Pentericci, Nonino, Fontanot, Menci, Monaco, Cristiani,
  {et~al.}}]{fontana2006}
Fontana, A., Salimbeni, S., Grazian, A., {et~al.} 2006, Astronomy \&
  Astrophysics, 459, 745

\bibitem[{{Ford} \& {McKernan}(2022)}]{Ford22}
{Ford}, K.~E.~S., \& {McKernan}, B. 2022, \mnras, 517, 5827,
  \dodoi{10.1093/mnras/stac2861}

\bibitem[{Fragione {et~al.}(2020)Fragione, Loeb, \& Rasio}]{Fragione_2020}
Fragione, G., Loeb, A., \& Rasio, F.~A. 2020, The Astrophysical Journal, 902,
  L26, \dodoi{10.3847/2041-8213/abbc0a}

\bibitem[{Fragos {et~al.}(2022)Fragos, Andrews, Bavera, Berry, Coughlin,
  Dotter, Giri, Kalogera, Katsaggelos, Kovlakas, Lalvani, Misra, Srivastava,
  Qin, Rocha, Roman-Garza, Serra, Stahle, Sun, Teng, Trajcevski, Tran, Xing,
  Zapartas, \& Zevin}]{posydon}
Fragos, T., Andrews, J.~J., Bavera, S.~S., {et~al.} 2022, POSYDON: A
  General-Purpose Population Synthesis Code with Detailed Binary-Evolution
  Simulations.
\newblock \doarXiv{2202.05892}

\bibitem[{Fryer {et~al.}(2012)Fryer, Belczynski, Wiktorowicz, Dominik,
  Kalogera, \& Holz}]{Fryer2012}
Fryer, C.~L., Belczynski, K., Wiktorowicz, G., {et~al.} 2012, The Astrophysical
  Journal, 749, 91, \dodoi{10.1088/0004-637x/749/1/91}

\bibitem[{{Furlong} {et~al.}(2015){Furlong}, {Bower}, {Theuns}, {Schaye},
  {Crain}, {Schaller}, {Dalla Vecchia}, {Frenk}, {McCarthy}, {Helly},
  {Jenkins}, \& {Rosas-Guevara}}]{Furlong2015}
{Furlong}, M., {Bower}, R.~G., {Theuns}, T., {et~al.} 2015, \mnras, 450, 4486,
  \dodoi{10.1093/mnras/stv852}

\bibitem[{{Gamba} {et~al.}(2023){Gamba}, {Breschi}, {Carullo}, {Albanesi},
  {Rettegno}, {Bernuzzi}, \& {Nagar}}]{Gamba2023Nature}
{Gamba}, R., {Breschi}, M., {Carullo}, G., {et~al.} 2023, Nature Astronomy, 7,
  11, \dodoi{10.1038/s41550-022-01813-w}

\bibitem[{Gangardt {et~al.}(2024)Gangardt, Trani, Bonnerot, \& Gerosa}]{pAGN}
Gangardt, D., Trani, A.~A., Bonnerot, C., \& Gerosa, D. 2024, Monthly Notices
  of the Royal Astronomical Society, 530, 3689, \dodoi{10.1093/mnras/stae1117}

\bibitem[{Gayathri {et~al.}(2022)Gayathri, Healy, Lange, O'Brien, Szczepanczyk,
  Bartos, Campanelli, Klimenko, Lousto, \& O'Shaughnessy}]{Gayathri2022}
Gayathri, V., Healy, J., Lange, J., {et~al.} 2022, Nature Astronomy, 6,
  344–349, \dodoi{https://doi.org/10.1038/s41550-021-01568-w}

\bibitem[{Hannam {et~al.}(2014)Hannam, Schmidt, Boh\'e, Haegel, Husa, Ohme,
  Pratten, \& P\"urrer}]{IMRPhenomPv2}
Hannam, M., Schmidt, P., Boh\'e, A., {et~al.} 2014, Phys. Rev. Lett., 113,
  151101, \dodoi{10.1103/PhysRevLett.113.151101}

\bibitem[{Harris {et~al.}(2020)Harris, Millman, van~der Walt, Gommers, \&
  et~al.}]{harris2020array}
Harris, C.~R., Millman, K.~J., van~der Walt, S.~J., Gommers, R., \& et~al.
  2020, Nature, 585, 357, \dodoi{10.1038/s41586-020-2649-2}

\bibitem[{Hunter(2007)}]{Hunter_2007}
Hunter, J.~D. 2007, Computing in Science \& Engineering, 9, 90,
  \dodoi{10.1109/MCSE.2007.55}

\bibitem[{Kritos {et~al.}(2024)Kritos, Strokov, Baibhav, \&
  Berti}]{Rapster2024}
Kritos, K., Strokov, V., Baibhav, V., \& Berti, E. 2024, Phys. Rev. D, 110,
  043023, \dodoi{10.1103/PhysRevD.110.043023}

\bibitem[{Li {et~al.}(2024{\natexlab{a}})Li, Tang, Gao, Wu, \&
  Wang}]{Li2024Field}
Li, Y.-J., Tang, S.-P., Gao, S.-J., Wu, D.-C., \& Wang, Y.-Z.
  2024{\natexlab{a}}, The Astrophysical Journal, 977, 67,
  \dodoi{10.3847/1538-4357/ad83b5}

\bibitem[{Li {et~al.}(2024{\natexlab{b}})Li, Wang, Tang, \&
  Fan}]{Li2024Origins}
Li, Y.-J., Wang, Y.-Z., Tang, S.-P., \& Fan, Y.-Z. 2024{\natexlab{b}}, Phys.
  Rev. Lett., 133, 051401, \dodoi{10.1103/PhysRevLett.133.051401}

\bibitem[{{LIGO Scientific Collaboration}(2018)}]{lalsuite}
{LIGO Scientific Collaboration}. 2018, {LIGO} {A}lgorithm {L}ibrary -
  {LALS}uite, free software (GPL), \dodoi{10.7935/GT1W-FZ16}

\bibitem[{Lyon {et~al.}(2024)Lyon, Cowley, Pye, \& Hopkins}]{Lyon2024}
Lyon, D.~J., Cowley, M.~J., Pye, O., \& Hopkins, A.~M. 2024, Decomposing
  Infrared Luminosity Functions into Star-Forming and AGN Components using
  CIGALE.
\newblock \doarXiv{2410.08541}

\bibitem[{Ma {et~al.}(2015)Ma, Hopkins, Faucher-Giguère, Zolman, Muratov,
  Kereš, \& Quataert}]{Ma15}
Ma, X., Hopkins, P.~F., Faucher-Giguère, C.-A., {et~al.} 2015, Monthly Notices
  of the Royal Astronomical Society, 456, 2140, \dodoi{10.1093/mnras/stv2659}

\bibitem[{Madau \& Fragos(2017)}]{Madau2017}
Madau, P., \& Fragos, T. 2017, The Astrophysical Journal, 840, 39,
  \dodoi{10.3847/1538-4357/aa6af9}

\bibitem[{{McKernan} {et~al.}(2022){McKernan}, {Ford}, {Callister}, {Farr},
  {O'Shaughnessy}, {Smith}, {Thrane}, \& {Vajpeyi}}]{qX22}
{McKernan}, B., {Ford}, K.~E.~S., {Callister}, T., {et~al.} 2022, \mnras, 514,
  3886, \dodoi{10.1093/mnras/stac1570}

\bibitem[{McKernan {et~al.}(2024)McKernan, Ford, Cook, Delfavero, Nathaniel,
  Postiglione, Ray, \& O'Shaughnessy}]{mcfactsI}
McKernan, B., Ford, K. E.~S., Cook, H.~E., {et~al.} 2024.
\newblock \doarXiv{2410.16515}

\bibitem[{{McKernan} {et~al.}(2014){McKernan}, {Ford}, {Kocsis}, {Lyra}, \&
  {Winter}}]{McK14}
{McKernan}, B., {Ford}, K.~E.~S., {Kocsis}, B., {Lyra}, W., \& {Winter}, L.~M.
  2014, \mnras, 441, 900, \dodoi{10.1093/mnras/stu553}

\bibitem[{{McKernan} {et~al.}(2012){McKernan}, {Ford}, {Lyra}, \&
  {Perets}}]{McK12}
{McKernan}, B., {Ford}, K.~E.~S., {Lyra}, W., \& {Perets}, H.~B. 2012, \mnras,
  425, 460, \dodoi{10.1111/j.1365-2966.2012.21486.x}

\bibitem[{{McKernan} {et~al.}(2020){McKernan}, {Ford}, {O'Shaugnessy}, \&
  {Wysocki}}]{McK20a}
{McKernan}, B., {Ford}, K.~E.~S., {O'Shaugnessy}, R., \& {Wysocki}, D. 2020,
  \mnras, 494, 1203, \dodoi{10.1093/mnras/staa740}

\bibitem[{{McKernan} {et~al.}(2018){McKernan}, {Ford}, {Bellovary}, {Leigh},
  {Haiman}, {Kocsis}, {Lyra}, {Mac Low}, {Metzger}, {O'Dowd}, {Endlich}, \&
  {Rosen}}]{McK18}
{McKernan}, B., {Ford}, K.~E.~S., {Bellovary}, J., {et~al.} 2018, \apj, 866,
  66, \dodoi{10.3847/1538-4357/aadae5}

\bibitem[{Miller \& Hamilton(2002)}]{Miller_2002}
Miller, M.~C., \& Hamilton, D.~P. 2002, The Astrophysical Journal, 576, 894,
  \dodoi{10.1086/341788}

\bibitem[{Morgan {et~al.}(2010)Morgan, Kochanek, Morgan, \& Falco}]{Morgan2010}
Morgan, C.~W., Kochanek, C.~S., Morgan, N.~D., \& Falco, E.~E. 2010, The
  Astrophysical Journal, 712, 1129, \dodoi{10.1088/0004-637X/712/2/1129}

\bibitem[{Morscher {et~al.}(2015)Morscher, Pattabiraman, Rodriguez, Rasio, \&
  Umbreit}]{Morscher_2015}
Morscher, M., Pattabiraman, B., Rodriguez, C., Rasio, F.~A., \& Umbreit, S.
  2015, The Astrophysical Journal, 800, 9, \dodoi{10.1088/0004-637x/800/1/9}

\bibitem[{Morton {et~al.}(2023)Morton, Rinaldi, Torres-Orjuela, Derdzinski,
  Vaccaro, \& Del~Pozzo}]{MortonGW190521}
Morton, S.~L., Rinaldi, S., Torres-Orjuela, A., {et~al.} 2023, Phys. Rev. D,
  108, 123039, \dodoi{10.1103/PhysRevD.108.123039}

\bibitem[{Nakajima {et~al.}(2023)Nakajima, Ouchi, Isobe, Harikane, Zhang, Ono,
  Umeda, \& Oguri}]{Nakajima_2023}
Nakajima, K., Ouchi, M., Isobe, Y., {et~al.} 2023, The Astrophysical Journal
  Supplement Series, 269, 33, \dodoi{10.3847/1538-4365/acd556}

\bibitem[{{Neumayer} {et~al.}(2020){Neumayer}, {Seth}, \&
  {B{\"o}ker}}]{Neumayer2020}
{Neumayer}, N., {Seth}, A., \& {B{\"o}ker}, T. 2020, \aapr, 28, 4,
  \dodoi{10.1007/s00159-020-00125-0}

\bibitem[{{Olejak} {et~al.}(2020){Olejak}, {Fishbach}, {Belczynski}, {Holz},
  {Lasota}, {Miller}, \& {Bulik}}]{Olejak2020}
{Olejak}, A., {Fishbach}, M., {Belczynski}, K., {et~al.} 2020, \apjl, 901, L39,
  \dodoi{10.3847/2041-8213/abb5b5}

\bibitem[{{Olejak, Aleksandra} {et~al.}(2024){Olejak, Aleksandra}, {Klencki,
  Jakub}, {Xu, Xiao-Tian}, {Wang, Chen}, {Belczynski, Krzysztof}, \& {Lasota,
  Jean-Pierre}}]{Olejak2024Spin}
{Olejak, Aleksandra}, {Klencki, Jakub}, {Xu, Xiao-Tian}, {et~al.} 2024, A\&A,
  689, A305, \dodoi{10.1051/0004-6361/202450480}

\bibitem[{Peng {et~al.}(2015)Peng, Maiolino, \& Cochrane}]{Peng2015}
Peng, Y., Maiolino, R., \& Cochrane, R. 2015, Nature, 521, 192

\bibitem[{Peters(1964)}]{Peters1964}
Peters, P.~C. 1964, Phys. Rev., 136, B1224, \dodoi{10.1103/PhysRev.136.B1224}

\bibitem[{Pordes {et~al.}(2007)Pordes, Petravick, Kramer, Olson, Livny, Roy,
  Avery, Blackburn, Wenaus, W{\"u}rthwein, Foster, Gardner, Wilde, Blatecky,
  McGee, \& Quick}]{osg07}
Pordes, R., Petravick, D., Kramer, B., {et~al.} 2007, in 78, Vol.~78, J. Phys.
  Conf. Ser., 012057, \dodoi{10.1088/1742-6596/78/1/012057}

\bibitem[{{Reines} {et~al.}(2013){Reines}, {Greene}, \& {Geha}}]{Reines2013}
{Reines}, A.~E., {Greene}, J.~E., \& {Geha}, M. 2013, \apj, 775, 116,
  \dodoi{10.1088/0004-637X/775/2/116}

\bibitem[{{Rodriguez} {et~al.}(2016){Rodriguez}, {Zevin}, {Pankow}, {Kalogera},
  \& {Rasio}}]{gwastro-popsynVclusters-Rodriguez2016}
{Rodriguez}, C.~L., {Zevin}, M., {Pankow}, C., {Kalogera}, V., \& {Rasio},
  F.~A. 2016, \apjl, 832, L2, \dodoi{10.3847/2041-8205/832/1/L2}

\bibitem[{Romero-Shaw {et~al.}(2020)Romero-Shaw, Lasky, Thrane, \&
  Bustillo}]{RomeroShawGW190521}
Romero-Shaw, I.~M., Lasky, P.~D., Thrane, E., \& Bustillo, J.~C. 2020, The
  Astrophysical Journal Letters, \dodoi{10.3847/2041-8213/abbe26}

\bibitem[{{Samsing} {et~al.}(2022){Samsing}, {Bartos}, {D'Orazio}, {Haiman},
  {Kocsis}, {Leigh}, {Liu}, {Pessah}, \& {Tagawa}}]{Samsing22}
{Samsing}, J., {Bartos}, I., {D'Orazio}, D.~J., {et~al.} 2022, \nat, 603, 237,
  \dodoi{10.1038/s41586-021-04333-1}

\bibitem[{{Santini} {et~al.}(2023){Santini}, {Gerosa}, {Cotesta}, \&
  {Berti}}]{Santini23}
{Santini}, A., {Gerosa}, D., {Cotesta}, R., \& {Berti}, E. 2023, \prd, 108,
  083033, \dodoi{10.1103/PhysRevD.108.083033}

\bibitem[{Schramm \& Silverman(2013)}]{Schramm_2013}
Schramm, M., \& Silverman, J.~D. 2013, The Astrophysical Journal, 767, 13,
  \dodoi{10.1088/0004-637X/767/1/13}

\bibitem[{{Secunda} {et~al.}(2019){Secunda}, {Bellovary}, {Mac Low}, {Ford},
  {McKernan}, {Leigh}, {Lyra}, \& {S{\'a}ndor}}]{Secunda19}
{Secunda}, A., {Bellovary}, J., {Mac Low}, M.-M., {et~al.} 2019, \apj, 878, 85,
  \dodoi{10.3847/1538-4357/ab20ca}

\bibitem[{{Secunda} {et~al.}(2020){Secunda}, {Bellovary}, {Mac Low}, {Ford},
  {McKernan}, {Leigh}, {Lyra}, {S{\'a}ndor}, \& {Adorno}}]{Secunda20}
---. 2020, \apj, 903, 133, \dodoi{10.3847/1538-4357/abbc1d}

\bibitem[{Sfiligoi {et~al.}(2009)Sfiligoi, Bradley, Holzman, Mhashilkar, Padhi,
  \& Wurthwein}]{osg09}
Sfiligoi, I., Bradley, D.~C., Holzman, B., {et~al.} 2009, in 2, Vol.~2, 2009
  WRI World Congress on Computer Science and Information Engineering, 428--432,
  \dodoi{10.1109/CSIE.2009.950}

\bibitem[{{Sirko} \& {Goodman}(2003)}]{SG03}
{Sirko}, E., \& {Goodman}, J. 2003, \mnras, 341, 501,
  \dodoi{10.1046/j.1365-8711.2003.06431.x}

\bibitem[{Stevenson \& Clarke(2022)}]{Stevenson2022}
Stevenson, S., \& Clarke, T.~A. 2022, Monthly Notices of the Royal Astronomical
  Society, 517, 4034

\bibitem[{{Stevenson} {et~al.}(2015){Stevenson}, {Ohme}, \&
  {Fairhurst}}]{Stevenson2015popsyn}
{Stevenson}, S., {Ohme}, F., \& {Fairhurst}, S. 2015, \apj, 810, 58,
  \dodoi{10.1088/0004-637X/810/1/58}

\bibitem[{{Stone} {et~al.}(2017){Stone}, {Metzger}, \& {Haiman}}]{Stone17}
{Stone}, N.~C., {Metzger}, B.~D., \& {Haiman}, Z. 2017, \mnras, 464, 946,
  \dodoi{10.1093/mnras/stw2260}

\bibitem[{{Tagawa} {et~al.}(2020{\natexlab{a}}){Tagawa}, {Haiman}, {Bartos}, \&
  {Kocsis}}]{Tagawa20b}
{Tagawa}, H., {Haiman}, Z., {Bartos}, I., \& {Kocsis}, B. 2020{\natexlab{a}},
  \apj, 899, 26, \dodoi{10.3847/1538-4357/aba2cc}

\bibitem[{{Tagawa} {et~al.}(2021){Tagawa}, {Haiman}, {Bartos}, {Kocsis}, \&
  {Omukai}}]{Tagawa21}
{Tagawa}, H., {Haiman}, Z., {Bartos}, I., {Kocsis}, B., \& {Omukai}, K. 2021,
  \mnras, 507, 3362, \dodoi{10.1093/mnras/stab2315}

\bibitem[{{Tagawa} {et~al.}(2020{\natexlab{b}}){Tagawa}, {Haiman}, \&
  {Kocsis}}]{Tagawa20}
{Tagawa}, H., {Haiman}, Z., \& {Kocsis}, B. 2020{\natexlab{b}}, \apj, 898, 25,
  \dodoi{10.3847/1538-4357/ab9b8c}

\bibitem[{{The LIGO Scientific Collaboration} {et~al.}(2020{\natexlab{a}}){The
  LIGO Scientific Collaboration}, Abbott, Abbott, \& Abbott}]{P1200087}
{The LIGO Scientific Collaboration}, Abbott, B.~P., Abbott, R., \& Abbott,
  T.~D. 2020{\natexlab{a}}, Living Reviews in Relativity, 23,
  \dodoi{10.1007/s41114-020-00026-9}

\bibitem[{{The LIGO Scientific Collaboration} \& {The Virgo
  Collaboration}(2021)}]{GWTC-2p1}
{The LIGO Scientific Collaboration}, \& {The Virgo Collaboration}. 2021,
  GWTC-2.1: Deep Extended Catalog of Compact Binary Coalescences Observed by
  LIGO and Virgo During the First Half of the Third Observing Run.
\newblock \doarXiv{2108.01045}

\bibitem[{{The LIGO Scientific Collaboration} {et~al.}(2020{\natexlab{b}}){The
  LIGO Scientific Collaboration}, {The Virgo Collaboration}, \& {The KAGRA
  Collaboration}}]{LIGO-O3-O3a-RP}
{The LIGO Scientific Collaboration}, {The Virgo Collaboration}, \& {The KAGRA
  Collaboration}. 2020{\natexlab{b}}, Available as LIGO-P2000077.
\newblock \url{https://dcc.ligo.org/LIGO-P2000077}

\bibitem[{{The LIGO Scientific Collaboration} {et~al.}(2021{\natexlab{a}}){The
  LIGO Scientific Collaboration}, {The Virgo Collaboration}, \& {The KAGRA
  Collaboration}}]{GWTC-3}
---. 2021{\natexlab{a}}, GWTC-3: Compact Binary Coalescences Observed by LIGO
  and Virgo During the Second Part of the Third Observing Run.
\newblock \doarXiv{2111.03606}

\bibitem[{{The LIGO Scientific Collaboration} {et~al.}(2021{\natexlab{b}}){The
  LIGO Scientific Collaboration}, {The Virgo Collaboration}, \& {The KAGRA
  Collaboration}}]{LIGO-O3-O3b-RP}
---. 2021{\natexlab{b}}, The population of merging compact binaries inferred
  using gravitational waves through GWTC-3.
\newblock \doarXiv{2111.03634}

\bibitem[{{Thompson} {et~al.}(2005){Thompson}, {Quataert}, \& {Murray}}]{TQM05}
{Thompson}, T.~A., {Quataert}, E., \& {Murray}, N. 2005, \apj, 630, 167,
  \dodoi{10.1086/431923}

\bibitem[{Tichy \& Marronetti(2008)}]{Tichy2008}
Tichy, W., \& Marronetti, P. 2008, Phys. Rev. D, 78, 081501,
  \dodoi{10.1103/PhysRevD.78.081501}

\bibitem[{{Ueda} {et~al.}(2014){Ueda}, {Akiyama}, {Hasinger}, {Miyaji}, \&
  {Watson}}]{Ueda2014}
{Ueda}, Y., {Akiyama}, M., {Hasinger}, G., {Miyaji}, T., \& {Watson}, M.~G.
  2014, \apj, 786, 104, \dodoi{10.1088/0004-637X/786/2/104}

\bibitem[{{Vajpeyi} {et~al.}(2022){Vajpeyi}, {Thrane}, {Smith}, {McKernan}, \&
  {Saavik Ford}}]{Vajpeyi22}
{Vajpeyi}, A., {Thrane}, E., {Smith}, R., {McKernan}, B., \& {Saavik Ford},
  K.~E. 2022, \apj, 931, 82, \dodoi{10.3847/1538-4357/ac6180}

\bibitem[{{Vink} {et~al.}(2001){Vink}, {de Koter}, \& {Lamers}}]{Vink2001}
{Vink}, J.~S., {de Koter}, A., \& {Lamers}, H.~J.~G.~L.~M. 2001, \aap, 369,
  574, \dodoi{10.1051/0004-6361:20010127}

\bibitem[{Virtanen {et~al.}(2020)Virtanen, Gommers, Oliphant, Haberland, Reddy,
  Cournapeau, Burovski, Peterson, Weckesser, Bright, {van der Walt}, Brett,
  Wilson, Millman, Mayorov, Nelson, Jones, Kern, Larson, Carey, Polat, Feng,
  Moore, {VanderPlas}, Laxalde, Perktold, Cimrman, Henriksen, Quintero, Harris,
  Archibald, Ribeiro, Pedregosa, {van Mulbregt}, \& {SciPy 1.0
  Contributors}}]{2020SciPy-NMeth}
Virtanen, P., Gommers, R., Oliphant, T.~E., {et~al.} 2020, Nature Methods, 17,
  261, \dodoi{10.1038/s41592-019-0686-2}

\bibitem[{Wang {et~al.}(2022)Wang, Li, Vink, Fan, Tang, Qin, \&
  Wei}]{Wang_2022}
Wang, Y.-Z., Li, Y.-J., Vink, J.~S., {et~al.} 2022, The Astrophysical Journal
  Letters, 941, L39, \dodoi{10.3847/2041-8213/aca89f}

\bibitem[{{Whitehead} {et~al.}(2024){Whitehead}, {Rowan}, {Boekholt}, \&
  {Kocsis}}]{Whitehead24}
{Whitehead}, H., {Rowan}, C., {Boekholt}, T., \& {Kocsis}, B. 2024, \mnras,
  531, 4656, \dodoi{10.1093/mnras/stae1430}

\bibitem[{{Wong} {et~al.}(2022){Wong}, {Breivik}, {Farr}, \&
  {Luger}}]{COSMICWong2022}
{Wong}, K. W.~K., {Breivik}, K., {Farr}, W.~M., \& {Luger}, R. 2022, arXiv
  e-prints, arXiv:2206.04062, \dodoi{10.48550/arXiv.2206.04062}

\bibitem[{Wysocki \& {O'Shaughnessy}(2018)}]{T1800427}
Wysocki, D., \& {O'Shaughnessy}, R. 2018, {Calibrating semi-analytic VTs
  against reweighted injection VTs} \url{https://dcc.ligo.org/LIGO-T1800427},
  LIGO Document Control Center.
\newblock \url{https://dcc.ligo.org/LIGO-T1800427}

\bibitem[{{Yang} {et~al.}(2019){Yang}, {Bartos}, {Gayathri}, {Ford}, {Haiman},
  {Klimenko}, {Kocsis}, {M{\'a}rka}, {M{\'a}rka}, {McKernan}, \&
  {O'Shaughnessy}}]{Yang19}
{Yang}, Y., {Bartos}, I., {Gayathri}, V., {et~al.} 2019, \prl, 123, 181101,
  \dodoi{10.1103/PhysRevLett.123.181101}

\bibitem[{{Zevin} {et~al.}(2020){Zevin}, {Spera}, {Berry}, \&
  {Kalogera}}]{Zevin2020}
{Zevin}, M., {Spera}, M., {Berry}, C. P.~L., \& {Kalogera}, V. 2020, \apjl,
  899, L1, \dodoi{10.3847/2041-8213/aba74e}

\bibitem[{{Zevin} {et~al.}(2021){Zevin}, {Bavera}, {Berry}, {Kalogera},
  {Fragos}, {Marchant}, {Rodriguez}, {Antonini}, {Holz}, \&
  {Pankow}}]{COSMICZevin2021}
{Zevin}, M., {Bavera}, S.~S., {Berry}, C. P.~L., {et~al.} 2021, \apj, 910, 152,
  \dodoi{10.3847/1538-4357/abe40e}

\bibitem[{Zwart {et~al.}(2004)Zwart, Baumgardt, Hut, Makino, \&
  McMillan}]{Portegies_Zwart_2004}
Zwart, S. F.~P., Baumgardt, H., Hut, P., Makino, J., \& McMillan, S. L.~W.
  2004, Nature, 428, 724, \dodoi{10.1038/nature02448}

\end{thebibliography}
\bibliographystyle{aasjournal}

\appendix
\section{Performance of \mcfacts{}}
\label{ap:performance}
\begin{table*}[!ht]
\begin{tabular}{|c|c|c|c|c|c|}
\hline
$\log_{10}(M_* [\msun])$ & $\log_{10}(M_{\mathrm{SMBH}} [\msun])$ & \SGfixed{}-early & \SGfixed-late & \SGscaled-early & \SGscaled-late \\
\hline
9.000  & 5.9292  & 1613.78 & 1267.38 & 1556.38 & 1258.69 \\
9.125  & 6.0692  & 1639.10 & 1479.97 & 1507.64 & 1643.80 \\
9.250  & 6.2092  & 1293.25 & 1247.75 & 1550.76 & 1349.67 \\
9.375  & 6.3492  & 1232.96 & 1899.62 & 1903.86 & 1877.29 \\
9.500  & 6.4892  & 1853.11 & 1955.90 & 1329.86 & 1757.00 \\
9.625  & 6.6292  & 1309.13 & 1914.80 & 1877.14 & 1291.35 \\
9.750  & 6.7692  & 1894.03 & 1242.61 & 1335.99 & 1937.42 \\
9.875  & 6.9092  & 1266.23 & 1842.99 & 1548.46 & 1464.62 \\
10.000 & 7.0492  & 1296.30 & 1948.08 & 1244.49 & 1298.54 \\
10.125 & 7.1892  & 1331.68 & 2140.83 & 1581.93 & 1823.61 \\
10.250 & 7.3292  & 1456.89 & 1557.39 & 1580.99 & 1799.17 \\
10.375 & 7.4692  & 1577.69 & 2677.19 & 1376.15 & 1578.56 \\
10.500 & 7.6092  &  -      & 2074.21 & 1748.60 & 1854.16 \\
10.625 & 7.7492  &  -      & 3463.55 & 1678.26 & 2394.42 \\
10.750 & 7.8892  &  -      &  -      & 1472.07 & 1809.36 \\
10.875 & 8.0292  &  -      &  -      & 1733.28 & 2162.29 \\
11.000 & 8.1692  &  -      & 4265.23 & 2145.17 & - \\
11.125 & 8.3092  &  -      & 3059.18 & 2877.01 & 5204.90 \\
11.250 & 8.4492  &  -      & 2763.61 & 3079.65 & 4771.14 \\
11.375 & 8.5892  &  -      &  -      & 2525.90 &  -      \\
11.500 & 8.7292  &  -      & 3815.73 & 2635.86 & 5098.95 \\
11.625 & 8.8692  &  -      & 2697.31 & -       & 4353.75 \\
11.750 & 9.0092  &  -      & 2726.99 & 3689.33 & 4253.54 \\
11.875 & 9.1492  &  -      & 2793.03 & -       & -       \\
12.000 & 9.2892  &  -      & 3477.99 & -       & -       \\
12.125 & 9.4292  &  -      &  -      & -       & -       \\
\hline
\end{tabular}
\caption{\label{tab:performance}
\textbf{Evaluation Time (s) of \mcfacts{} for different disk models:}
The \texttt{perf\_counter} time for the evaluation of 100 galaxy realizations
    using \mcfacts{} with the various disk models outlined in Section \ref{sec:disk}
    for 10 Myr.
These reference times were recorded using
    a single ARM64 core with a clock speed of 3.0 GHz.
    Missing entries represent runs which were completed on a different computer.
}
\end{table*}
 Table \ref{tab:performance} records the evaluation time for \mcfacts{},
    for the disk models identified in Section \ref{sec:disk}.

\end{document}